\documentclass
[superscriptaddress,secnumarabic,amssymb,amsmath,nobibnotes,aps,prd,showpacs,nofootinbib]{revtex4}%
\usepackage{graphicx}
\usepackage{epsf}
\usepackage{bm}
\usepackage{amsmath}
\usepackage{amsfonts}
\usepackage{amssymb}
\usepackage{epstopdf}
\usepackage{natbib}
\usepackage{color}%
\providecommand{\U}[1]{\protect\rule{.1in}{.1in}}
\newcommand{\be}{\begin{equation}}
\newcommand{\ee}{\end{equation}}

\newcommand{\mincir}{\raise
-3.truept\hbox{\rlap{\hbox{$\sim$}}\raise4.truept\hbox{$<$}\ }}
\newcommand{\magcir}{\raise
-3.truept\hbox{\rlap{\hbox{$\sim$}}\raise4.truept\hbox{$>$}\ }}

\begin{document}
\title{Cosmological solutions with gravitational particle production and non-zero curvature}
\author{Andronikos Paliathanasis}
\email{anpaliat@phys.uoa.gr}
\affiliation{Instituto de Ciencias F\'{\i}sicas y Matem\'{a}ticas, Universidad Austral de
Chile, Valdivia, Chile}
\author{John D. Barrow}
\email{jdb34@damtp.cam.ac.uk}
\affiliation{DAMTP, Centre for Mathematical Sciences, University of Cambridge, Cambridge
CB3 0WA, U.K.}
\author{Supriya Pan}
\email{span@iiserkol.ac.in}
\affiliation{Department of Physical Sciences, Indian Institute of Science Education and
Research -- Kolkata, Mohanpur -- 741246, West Bengal, India}

\pacs{98.80.-k, 05.70.Ln, 04.40.Nr, 98.80.Cq.}

\begin{abstract}
In a homogeneous and isotropic universe with non-zero spatial curvature we
consider the effects of gravitational particle production in the dynamics of
the universe. We show that the dynamics of the universe in such a background
is characterized by a single nonlinear differential equation which is
significantly dependent on the rate of particle creation and whose solutions
can be dominated by the curvature effects at early times. For different
particle creation rates we apply the singularity test in order to find the
analytic solutions of the background dynamics. We describe the behavior of the
cosmological solutions for both open and closed universes. We also show how
the effects of curvature can be produced by the presence of a second perfect
fluid with an appropriate equation of state. {By combining that results
with the analysis of the critical points we find that our consideration can be
related with the pre-inflationary era. Specifically we find that for negative
spatial curvature small changes of the Milne spacetime leads to a de Sitter
universe.}

\end{abstract}
\maketitle




\section{Introduction}

In this work we investigate the effects of particle production on the early
stages and later evolution of cosmological models with non-zero spatial
curvature. Unlike in standard perfect fluid cosmologies without particle
production or viscous effects, the curvature terms dominate at very early times.

Cosmological particle production has been intensively examined as part of the
universe's early evolution. The first investigations focussed on the
potentially isotropizing effects of created particles in anisotropic universes
\cite{aniso}, although they ran the risk of overproducing entropy
\cite{BMatz}. Grishchuk \cite{grish} provided a classical counterpart to
quantum production that produced amplification of gravitational waves when the
expansion dynamics deviated from those of radiation-domination. Later, many
investigations showed how particle production by a time-varying gravitational
field in an early inflationary era of accelerated expansion could generate
observable effects in the form of scalar and tensor perturbation modes. A
period of late-time accelerated expansion might also be described in such a
framework, see for example refs. \cite{SSL09, LJO10, LB10, JOBL11, LBC12,
LGPB14, Ramos:2014dba, CPS14, FPP14, NP15, lima16, NP16, HP16, PHPS16}. This
particle production mechanism offers an alternative way to view the early
accelerating phase \cite{Prigogine-inf, Abramo:1996ip, Gunzig:1997tk,
Zimdahl:1999tn, RCN16} and it has also been noticed that it allows for a
universe which experiences two de Sitter phases (at early and late times) with
an intermediate period of radiation and matter dominated phases \cite{LBC12,
PHPS16}. Moreover, gravitational particle production may have some impact on
the structure formation in our universe \cite{RCN16a}. Originally, research in
this direction was motivated by the macroscopic particle production at the
expense of gravitational field in a background of a non-equilibrium
thermodynamics which naturally incorporates a negative back-reaction term in
the Einstein's field equations \cite{Prigogine88} and hence, without any
presumption of a dark energy component or modifications to gravity, an
accelerating universe can be realized. Just as the equation of state for any
dark energy fluid, or the particular forms for any modification of general
relativity play a key role in determining different cosmic phases, so with
particle production the dominant role is played by the rate of particle
production, $\Gamma$ \cite{SSL09, LJO10, LB10, JOBL11, LBC12, LGPB14,
Ramos:2014dba, CPS14, FPP14, NP15, lima16, NP16, HP16, PHPS16, Abramo:1996ip,
Gunzig:1997tk, Zimdahl:1999tn, RCN16}.

In what follows we investigate the cosmological solutions with `adiabatic'
particle production in the presence of spatial curvature (or an equivalent
`fluid'). However, it is well known that non-zero curvature effects must be
very small today to be in agreement with some observational data
\cite{Ade:2015xua, DiDio:2016ykq}. This approximate spatial flatness of the
late-time universe is a prediction of most inflationary cosmologies
\cite{Guth:2013sya,Adler:2005mn,Kleban:2012ph,Guth:2012ww,Lake:2004xg}. It
corresponds to $\Omega\equiv\rho/\rho_{c}=1\pm\delta$, where $\rho$ is the
total energy density of the universe and $\rho_{c}$ denotes the energy density
for the flat universe, and $\delta$ is at least as large as the magnitude of
any density perturbations created over large scales by inflation. Now, in any
cosmological model that seeks to describe the evolution from an early
inflationary era to the current accelerating state, it might leave a small
deviation from spatial flatness after the inflationary phase. In particular,
predictions of non-zero mean curvature, which are less than or equal to the
level of the curvature fluctuations ($\Omega_{k}\lesssim10^{-5}$) have little
meaning. On the other hand, it is quite interesting to mention that the
inflation is also possible with $\Omega>1$ \cite{Linde:2003hc}, or $\Omega<1$
\cite{Gott,Kamionkowski:1993cv,Kamionkowski:1994sv,Linde1996,Bucher:1994gb,Linde:1998gs}%
. The observational consequences of the curved inflationary models have been
investigated recently, \cite{DiDio:2016ykq,Bull:2013fga, Bonga:2016iuf}, with
a conclusion that such models may be compatible with the observational data.
In particular, in \cite{steingman}, it was shown that inflation is immune to
negative curvature while the energy density which corresponds to the curvature
can be nonzero in the pre-inflationary era \cite{asla1,asla2}. So, in
principle, it is possible to construct homogeneous and isotropic open or
closed inflationary scenarios that would lead to a spatially flat universe at
present time. Future observations may narrow these possibilities further.
Thus, in the present work we include spatial curvature to study the theory of
particle production.

In this work we take the particle production rates to be free parameters and
start from the simplest possibility, of a constant creation rate, and then
progress to a very generalized functional form. This is motivated by the
discovery that the presence of spatial curvature can dominate the dynamics
near the singularity in analogous bulk viscous cosmology \cite{JDB1988}: this
is the opposite to the situation without bulk viscosity or particle creation
present. We focus mainly on the exact analytic solutions of the Einstein's
field equations by applying the method of singularity analysis to the
differential equations as we did recently in modified gravity theories
\cite{Paliathanasis:2016tch, Paliathanasis:2016vsw}. We find that the
gravitational equations for two specific models of matter creation pass the
singularity test, which means that the solution can be written as a Laurent
series around the movable singularity. We discuss the possible evolution of
open and closed universes driven by gravitational particle production.

The plan of the paper is as follows. In Section \ref{sec:ede} we define our
cosmological model with matter creation terms. In order to investigate the
existence of analytical solutions, we apply the method of movable
singularities to the differential equations for two models. For arbitrary
matter creation rate, we find that the scale factor $a\left(  t\right)  \simeq
t$ satisfies the field equations. In Section \ref{appendi} \ we study the
existence of de Sitter points for the field equations by performing a
dynamical analysis of the critical points. In Section \ref{sec:constantRate}
we give the exact solution for a constant matter creation rate, while in
Section \ref{sec:varyingI} we study the time-varying particle creation
functions $\Gamma_{A}\propto1/H~$and $\Gamma_{B}\propto H^{2}$ . The stability
properties of the special solution $a\left(  t\right)  \simeq t$ are also
studied. In Section \ref{sec:varyingII}, we consider a general time-varying
particle production rate and we present the corresponding cosmological
solutions at the small and large time limits. Further, in section
\ref{two-fluid}, we have briefly outlined how the effect of a non zero
curvature in the cosmology of matter creation is recovered when a second
perfect fluid is introduced as an alternative to the curvature fluid. Finally,
in Section \ref{discu} we discuss our results and draw conclusions.

\section{Basic model of cosmological particle production}

\label{sec:ede}

We assume that the geometry of our universe is described by the
Friedmann-Lema\^{\i}tre-Robertson-Walker (FLRW) metric
\begin{equation}
ds^{2}=-dt^{2}+a^{2}(t)\left[  \frac{dr^{2}}{(1-kr^{2})}+r^{2}\left(
d\theta^{2}+\sin^{2}\theta\,d\phi^{2}\right)  \right]  , \label{cp.01}%
\end{equation}
where $a(t)$ is the expansion scale factor, and $k$ is the curvature scalar
which describes a flat, open or closed universe for $k=0,\,-1,\,+1$,
respectively. In such a metric, the Einstein's field equations for a fluid
endowed with particle creation can be written as%

\begin{equation}
\frac{\dot{a}^{2}}{a^{2}}+\frac{k}{a^{2}}=\frac{\rho}{3}, \label{friedmann1}%
\end{equation}%
\begin{equation}
2\,\frac{\ddot{a}}{a}+\frac{\dot{a}^{2}}{a^{2}}+\frac{k}{a^{2}}=-(p+p_{c}),
\label{friedmann2}%
\end{equation}
where an overdot represents one time cosmic time differentiation; $\rho$, $p$,
are respectively the energy density and the thermodynamic pressure of the
matter content; and $p_{c}$ denotes the creation pressure which is related to
the gravitationally induced `adiabatic' particle production by \cite{SSL09,
LJO10, LB10, JOBL11, LBC12, LGPB14, Ramos:2014dba, CPS14, FPP14, NP15, lima16,
NP16, HP16, PHPS16}%

\begin{equation}
p_{c}=-\frac{\Gamma}{3H}(p+\rho)\,, \label{cp}%
\end{equation}
where $H=\dot{a}/a$ is the Hubble rate of the FLRW universe, and $\Gamma(t)$
is some unknown creation rate. In principle, the exact functional form of
$\Gamma$ can be determined from the quantum field theory in curved spacetimes.
However, to close the system of equations (\ref{friedmann1}) and
(\ref{friedmann2}), we assume the fluid component obeys a perfect fluid
equation of state $p=(\gamma-1)\rho$, where $\gamma$ is a constant lying in
$\left(  0,2\right)  $. So, $\gamma=4/3$, i.e. $p=\frac{\rho}{3}$, denotes a
radiation-like fluid; $\gamma=1$, equivalently, $p=0$, is a pressureless dark
matter-like fluid; and $\gamma=0$, describes a cosmological constant-like
term, i.e. $p=-\rho$. Now, the question arises what kind of particles are
being created by the gravitational field. Until now we do not have any exact
explanation on the created particles. Therefore, we have assumed that the
created particles are simply the perfect fluid particles but with unknown
nature. However, there are some arguments in the current literature following
the local gravity constraints \cite{lgc1, lgc2, lgc3} which claim that the
production of baryonic particles at present universe are much limited, and the
radiation has practically no impact on the current dynamics of the universe.
On the other hand, during very early evolution of the universe, when the
universe was dominated by hot radiation, according to Parker's theorem, the
production of massless particles or photon or radiation is strongly suppressed
\cite{PT}, that means the creation rate is almost zero but not equal to zero,
at least from theoretical grounds. Thus, based on these consequences, just for
simplicity one often argues that the produced particles are the dark matter
particles. Nevertheless, we would like to remark that one can introduce the
basic thermodynamic variables into the equation of state of the fluid, for
instance the temperature, $T$ and the number density, $n$ of the particles as
$p=p(n,T)$, and $\rho=\rho(n,T)$ \cite{Zimdahl:1996ka}. Specifically, an ideal
gas type equation of state of the fluid given by $(\rho,p)=(mn+\gamma nT$,
$nT$) where $m$ is the mass of each particle, and $\gamma$ is a constant.
However, we must be cautious. The equation of state of matter in the universe
at temperatures approaching the Planck scale is difficult to assess because
the rate of asymptotically-free interactions is slower than the Hubble
expansion rate when $T>10^{15}GeV$. There is no equilibrium, or `temperature',
in the usual sense because of this. As the Planck scale is approached the
statistical basis of statistical mechanics is questionable because of the
small number of particles inside the particle horizon (unless $a=t$).
Moreover, the ideal gas approximation may fail. Typically, it requires the
mean free path between interactions to exceed the interparticle spacing in the
early universe; aside from at phase transitions, this requires the product of
the number of effectively relativistic spin states, $g_{\ast}$,$\ $and the
dimensionless interaction strength $\alpha(T)$ to satisfy $g_{\ast}%
^{1/3}\alpha\lesssim10$, and this can easily fail at energies well below the
Planck energy of $10^{19}GeV$. Therefore, we consider a wide range of
effective behaviours and look for features that do not depend sensitively on
our choice of the effective thermodynamic parameters. With regard to the
equation of state parameter for the perfect fluid, we assume that $p=\left(
\gamma-1\right)  \rho$ with $\gamma$ constant. Indeed, the equation of state
parameter of the perfect gas is the simplest that can be chosen but it is
merely an ansatz since the effective equation of state parameter of the very
early universe is unknown. However, the particle creation term provides an
effective pressure term which modifies the equation of state parameter away
from a perfect fluid. In an isotropic universe we know that a bulk viscosity
is the only form of non-equilibrium behaviour allowed and eq. (\ref{cp}) is a
particular expression of it. It reflects the response of the cosmological
fluid to being pulled out of equilibrium by the expansion of the universe and
is the classical model of quantum particle production.

Now, for the perfect fluid equation of state with the production term
(\ref{cp}), the essential conservation equation ($T_{;\nu}^{\mu\nu}=0$) can be
written and solved as%

\begin{align}
\dot{\rho}+3H\left(  1-\frac{\Gamma}{3H}\right)  (p+\rho)  &
=0\,\Longleftrightarrow\,\dot{\rho}+3\gamma\,\frac{\dot{a}}{a}\,\rho
=\gamma\Gamma\rho,\label{conservation}\\
\rho &  =\frac{A}{a^{3\gamma}}\exp[\gamma\int\Gamma(t)dt], \label{consol1}%
\end{align}
which follows from (\ref{friedmann1}) and (\ref{friedmann2}). \ Recall that,
for a perfect fluid with time-varying equation of state parameter $w\left(
a\right)  $, the conservation equation gives that
\[
\rho=\exp\left[  \int3\left(  1+w\left(  a\right)  \right)  d\ln a\right]
\]
which means that the particle creation rate $\Gamma$ is related to the
time-varying equation of state parameter $w(a)$. Hence the same $\Gamma$
provides different effective fluids, and consequently different gravitational
effects, for the same $\gamma$. On the other hand the idea of the particle
creation approach is to modify the effective equation of state parameter of
the perfect fluid.

For slow particle production, $\Gamma\ll3H$, eqn. (\ref{conservation}) reduces
to $\dot{\rho}+3H(p+\rho)=0$, and $\rho\propto a^{-3\gamma}$. This means that
the standard evolution equation of the perfect fluid is recovered. However,
$\Gamma>0$ can allow the exponential particle production term to dominate the
adiabatic decay term in (\ref{consol1}). So, clearly, the rate of particle
production significantly affects the evolution equations. One may find that a
nonvanishing creation rate, $\Gamma$, is similar dynamically to the effects of
a bulk viscous pressure \cite{Zeldovich70,Hu1982,JDB1986, JDB1987, JDB1988,
JDB1990}, although they differ from the thermodynamic view point
\cite{Lima:1992np}. The right-hand side of eqn. (\ref{conservation}) is
replaced by $9H^{2}\eta$ in the presence of a bulk viscosity coefficient
$\eta$, which is often chosen to be proportional to a power of $\rho$
\cite{JDB1988}.

An important question immediately follows from these observations as to what
the possible forms of $\Gamma$ should be. The particle creation process is
motivated by the quantum field theory (QFT) in curved spacetime \cite{Birrell}%
, so in principle one can determine $\Gamma$ using QFT in curved spacetime.
Unfortunately, QFT cannot yet provide an exact functional form for $\Gamma$.
Therefore, it is more convenient to make a general choice of the rate and
examine its effects on the cosmological model. This will allow us to determine
constraints on the rate of particle production from its testable effects on
cosmological evolution. By combining any two independent field equations of
(\ref{friedmann1}), (\ref{friedmann2}) and (\ref{conservation}), we get the
following single second-order nonlinear differential equation%

\begin{equation}
2\,\dot{a}\,\ddot{a}+(3\gamma-2)\,\frac{\dot{a}}{a}\,\left(  \dot{a}%
^{2}+k\right)  -\Gamma\,\gamma\left(  \dot{a}^{2}+k\right)  =0, \label{ode}%
\end{equation}
whose solution tells us the evolution of $a(t)$ once the creation rate,
$\Gamma(t)$, is specified. It is easy to see that the equation (\ref{ode})
always admits the power-law solution,%
\begin{equation}
a\left(  t\right)  =a_{0}\,t, \label{sol1}%
\end{equation}
where $\left(  a_{0}\right)  ^{2}=-k$, which means that the signature of the
spacetime is Lorentzian if ~$k<0$. It is important to note that, for the
solution $a(t)=a_{0}\,t$ in Eq. (\ref{sol1}), we have $\rho=0$, hence $p=0$
and so from Eq. (\ref{cp}) we find that $p_{c}=0$. This means that we do not
experience any effects of matter creation in the cosmological solutions. Let
us introduce the total equation of state of the fluid, $w_{tot}\equiv P/\rho$,
where $P=p+p_{c},$ with%

\begin{equation}
w_{tot}=-1-\frac{2}{3}\left(  \frac{\dot{H}-k/a^{2}}{H^{2}+k/a^{2}}\right)
=-1+\gamma\left(  1-\frac{\Gamma}{3H}\right)  .
\end{equation}
This means that the total fluid can exhibit different scenarios depending on
the particle creation rate $\Gamma$, as described in \cite{HP16}.
Specifically, when $\Gamma\ll3H$, $w_{tot}\simeq(-1+\gamma)$ $>-1$, for
$\gamma>0$ ; when $\Gamma\gg3H$, $w_{tot}<-1$ for $\gamma>0$; and when
$\Gamma=3H$, $w_{tot}=-1$, irrespective of the sign in $\gamma$. The
interesting thing is that one can get a phantom dominated universe by the
introduction of particle creation without invoking any phantom scalar field
\cite{caldwell}.


\section{Dynamical point of view}

\label{appendi}

There are various ways to study the evolution of the cosmological models. One
of the main methods is the analysis of the critical points. This can provide
us with information about the different phases in the evolution of the universe.

Let us introduce the new variable $\Omega=\frac{\rho}{3H^{2}}$, which
corresponds to the Hubble-normalised energy density of the matter source. In
these new variables the field equations (\ref{friedmann2}%
),\ (\ref{conservation}), can be written as:%
\begin{equation}
\dot{H}=\left(  2-3\gamma+\frac{\gamma\Gamma\left(  H\right)  }{H}\right)
\Omega H^{2}-H^{2},\label{eq.01}%
\end{equation}%
\begin{equation}
\dot{\Omega}=\Omega\left(  2\Omega-1\right)  \left[  \left(  3\gamma-2\right)
H-\gamma\Gamma\left(  H\right)  \right]  ,\label{eq.02}%
\end{equation}
while the constraint equation (\ref{friedmann1}) becomes%
\begin{equation}
1-\Omega+\frac{k}{a^{2}H^{2}}=0.\label{eq.03}%
\end{equation}

Every critical (fixed) point $P=\left(  H\left(  P\right)  ,\Omega\left(
P\right)  \right)  $ of the dynamical system (\ref{eq.01}), (\ref{eq.02})
corresponds to a state in which the Hubble function is constant, i.e.
$H=H_{P}$, and consequently the solution of the scale factor is $a\left(
t\right)  \simeq e^{H_{P}t},~$with $H_{P}\neq0$. That means that the critical
points of (\ref{eq.01}), (\ref{eq.02}) describe de Sitter solutions. In
addition, the constraint equation (\ref{eq.03}) must be satisfied in order
that these points exist. \ \
As mentioned in the introduction, in the following subsections we will
concentrate on the cases with three representative particle-creation rates:
$\Gamma\left(  H\right)  =\Gamma_{0}$, $\Gamma_{A}\left(  H\right)
=\Gamma_{0}+\frac{\Gamma_{1}}{H}$, and $\Gamma_{B}\left(  H\right)
=\Gamma_{0}+\Gamma_{1}H^{2}$.

\subsection{Constant creation rate: $\Gamma=\Gamma_{0}$}

In the early universe when $H$ is very large, the quantity $\Gamma_{0}/3H$
becomes very small and we recover the standard evolutionary law for the case
of a perfect fluid; but, at late times, a constant creation rate creates a
significant deviation from the standard cosmological laws. However, note that
the solution of the master equation (\ref{ode}) with $k=0$, still predicts an
initial big bang singularity \cite{HP16}.

For a constant creation rate, and for $\gamma\in\left(  0,2\right)  $, we find
that the system (\ref{eq.01}), (\ref{eq.02}) has the critical point
$P_{0}=\left(  H\left(  P_{0}\right)  ,\Omega\left(  P_{0}\right)  \right)  ,$
with coordinates $P_{0}=\left(  \frac{\Gamma_{0}}{3},\frac{1}{2}\right)  $.
Moreover, from (\ref{eq.03}), we find that $18k+a^{2}\Gamma_{0}^{2}=0$, which
means that the point exists only for negative $k.$ The eigenvalues of the
linearized system around the point $P_{0}$ are calculated to be $e_{1}%
=-\frac{2}{3}\Gamma_{0}~,~e_{2}=-\frac{1}{2}\gamma\Gamma_{0}$, which means
that the point is an attractor for $\Gamma_{0}>0$.

\subsection{Creation rate: $\Gamma\propto1/H~$and $\Gamma\propto H^{2}$}

The two creation rates $\Gamma_{A}\left(  H\right)  $ and $\Gamma_{B}\left(
H\right)  $ have been studied previously for the case of a spatially flat
universe\footnote{If we assume the time-varying matter creation rate to be
$\Gamma\left(  H\right)  =\bar{\Gamma}\left(  H\right)  +\Gamma_{1}H$, then
the linear term does not alter the dynamics, since for another value of
$\gamma$, equation (\ref{ode}) will have that same form.}, where it was found
that exponential expansion can occur \cite{Prigogine-inf, Abramo:1996ip,
Gunzig:1997tk}. On the other hand, the matter creation rate $\Gamma_{A}\left(
H\right)  $ \ in a spatially flat universe creates accelerating cosmological
solutions \cite{pan-ahep} in which the expansion approaches an asymptotic de
Sitter stage. Furthermore, if the created particles are dark matter-like with
$\Gamma_{A}\propto1/H$, then in the flat FLRW universe we get a mirage of
$\Lambda$CDM cosmology known as the creation-cold-dark-matter (CCDM) cosmology
\cite{Ramos:2014dba}. On the other hand, the second creation rate $\Gamma_{B}$
can be rewritten as $\Gamma_{B}=\Gamma_{0}+\Gamma_{1}\,(\rho/3)$, so the
global behavior of the universe is driven by matter creation.

Therefore, for the creation rate $\Gamma_{A}\left(  H\right)  ,$ we find that
the field equations (\ref{eq.01}), (\ref{eq.02}) provide us with two de Sitter
solutions, points $P_{+}^{A}$, and $P_{-}^{A}$ with coordinates~$P_{\pm}%
^{A}=\left(  \frac{\Gamma_{0}\pm\sqrt{\left(  \Gamma_{0}\right)  ^{2}%
+12\Gamma_{1}}}{6},\frac{1}{2}\right)  ;$ while the constraint (\ref{eq.03})
gives $1-72k\left(  a\left(  H\left(  P_{\pm}^{A}\right)  \right)  \right)
^{-2}=0$. This means that the points exist only for negative spatial
curvature. \ With regard to the stability of the points, we find that
$P_{-}^{A}$ is always unstable for $\gamma\in\left(  0,2\right)  $, while
point~$P_{+}^{A}$ is always stable for $\Gamma_{1}>0$.

Furthermore, for the third creation rate in our study, namely $\Gamma
_{B}\left(  H\right)  $, the field equations (\ref{eq.01}) and (\ref{eq.02})
again admit two critical points, $P_{+}^{B}$, and $P_{-}^{B}$, with
coordinates~$P_{\pm}^{B}=\left(  \frac{3\pm\sqrt{9-4\Gamma_{0}\Gamma_{1}}%
}{2\Gamma_{1}},\frac{1}{2}\right)  $ if and only if $k$ is negative, because
it is given by (\ref{eq.03}). Finally, the point $P_{+}^{B}$ is always
unstable, and $P_{-}^{B}$ describes an attractor when $\Gamma_{1}<0$.

\begin{figure}[t]
\centering\includegraphics[scale=0.45]{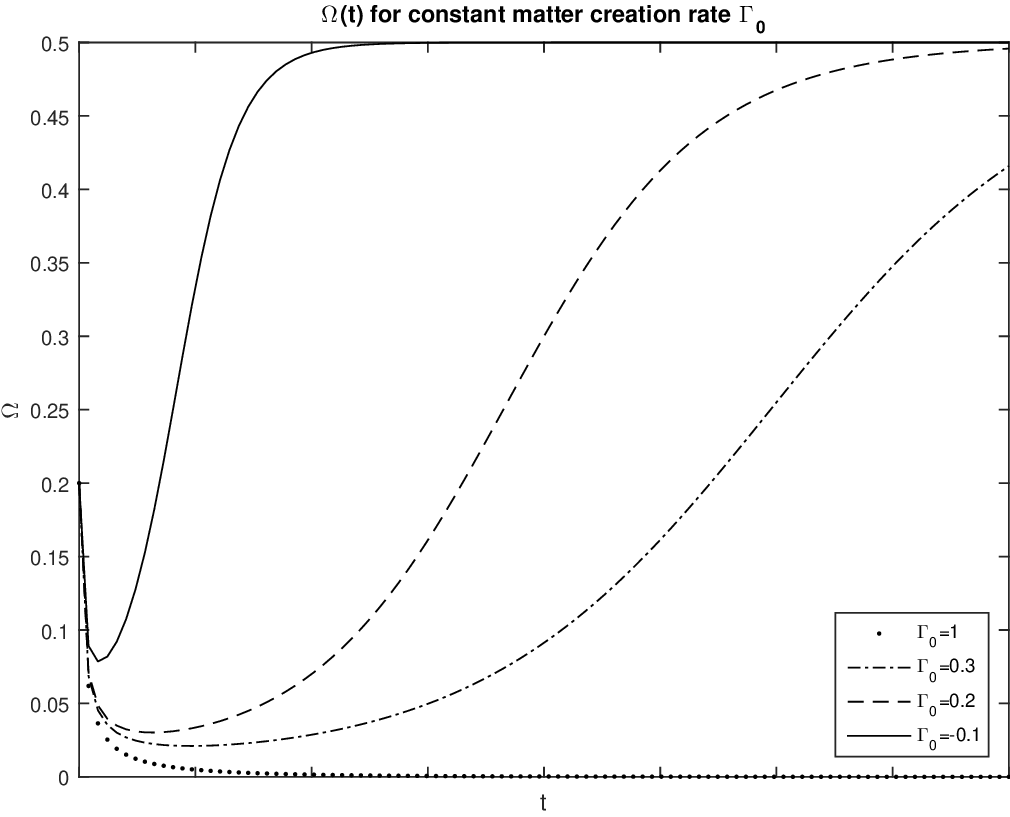}
\centering\includegraphics[scale=0.45]{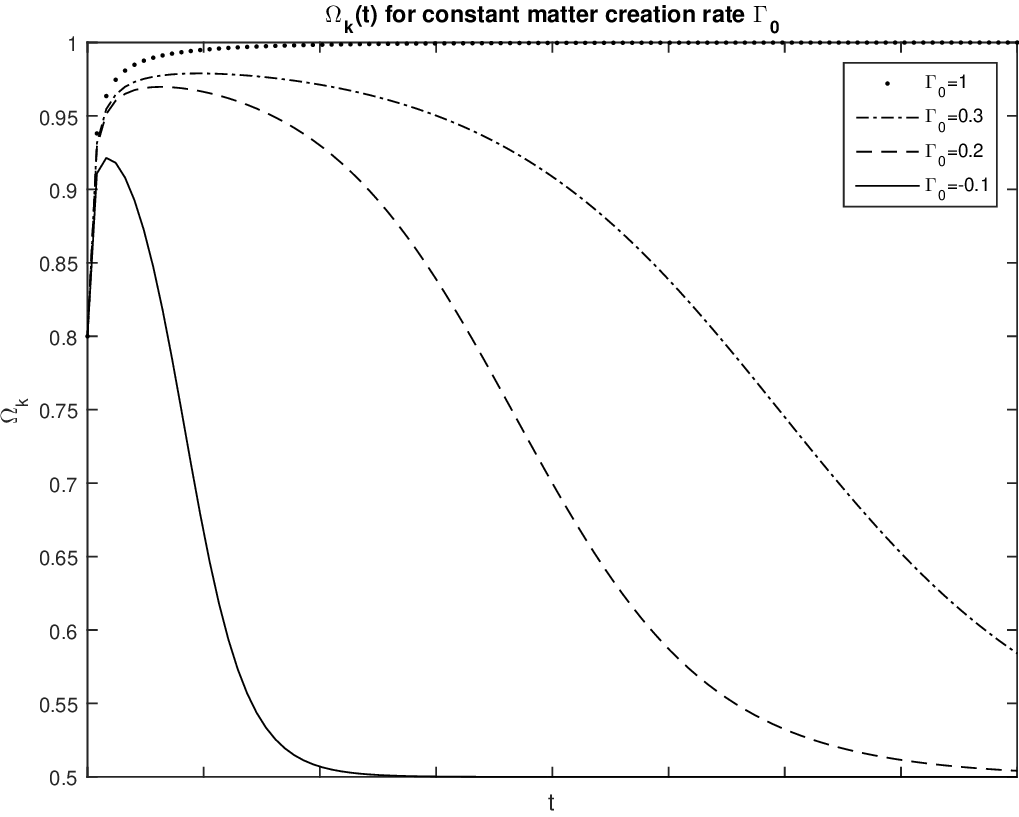}
\caption{Evolution of the energy density $\Omega$ (left fig.) and $\Omega
_{k}~$(right fig.) for the constant matter creation rate $\Gamma_{0}~$and
negative spatial curvature, $k=-1$. The lines are for $\gamma=1$, i.e. dust
fluid. Solid line is for $\Gamma_{0}=1,$ \ dash-dash line for $\Gamma
_{0}=0.3,$ dash-dot line for $\Gamma_{0}=0.2$ and dot-dot line for $\Gamma
_{0}=-0.1.$}%

\label{omegag01}%
\end{figure}

\begin{figure}[t]
\centering\includegraphics[scale=0.45]{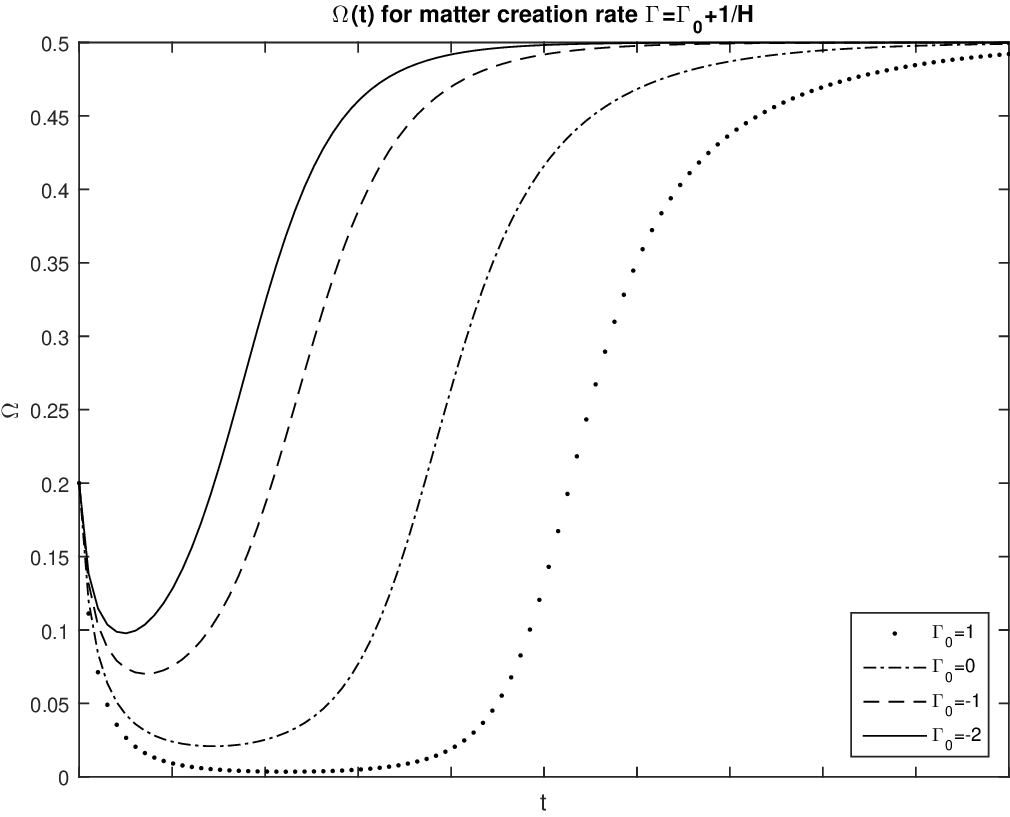}
\centering\includegraphics[scale=0.45]{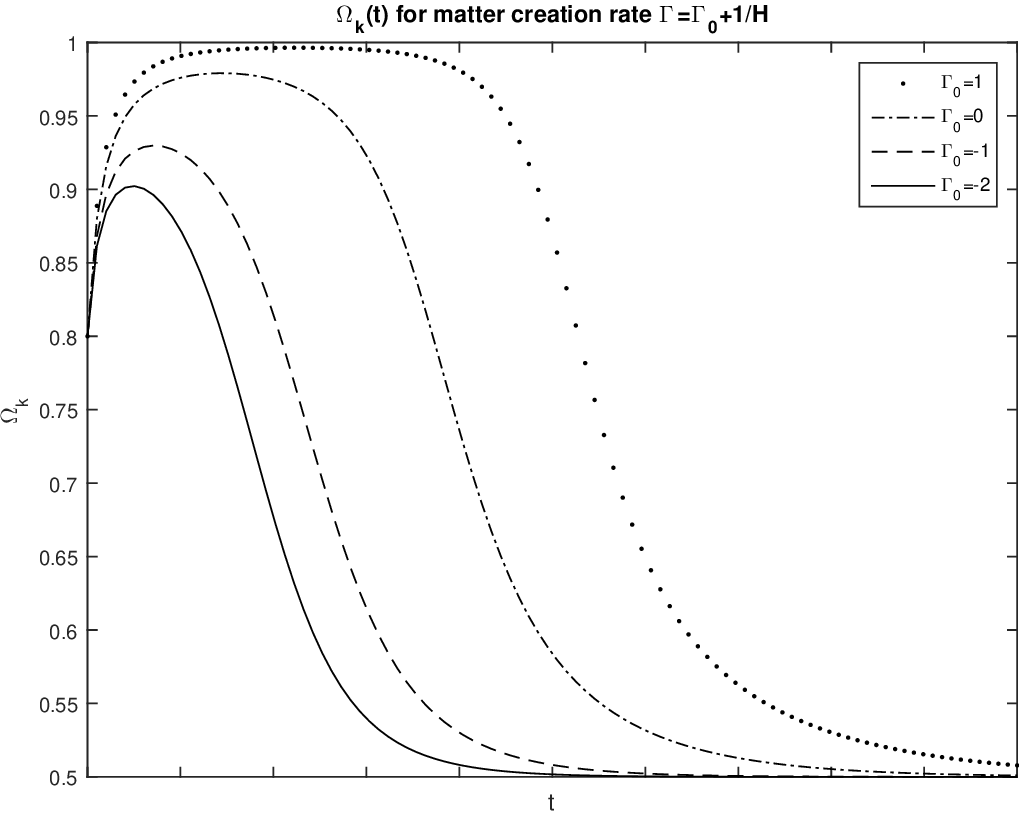}
\centering\includegraphics[scale=0.45]{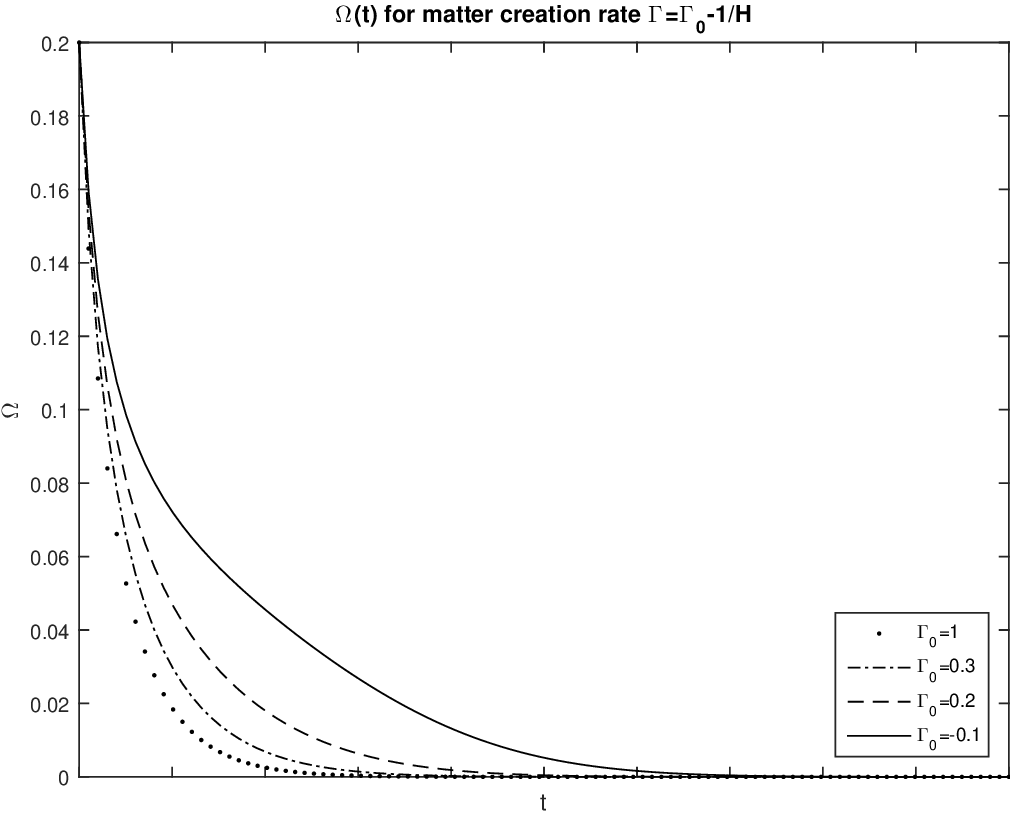}
\centering\includegraphics[scale=0.45]{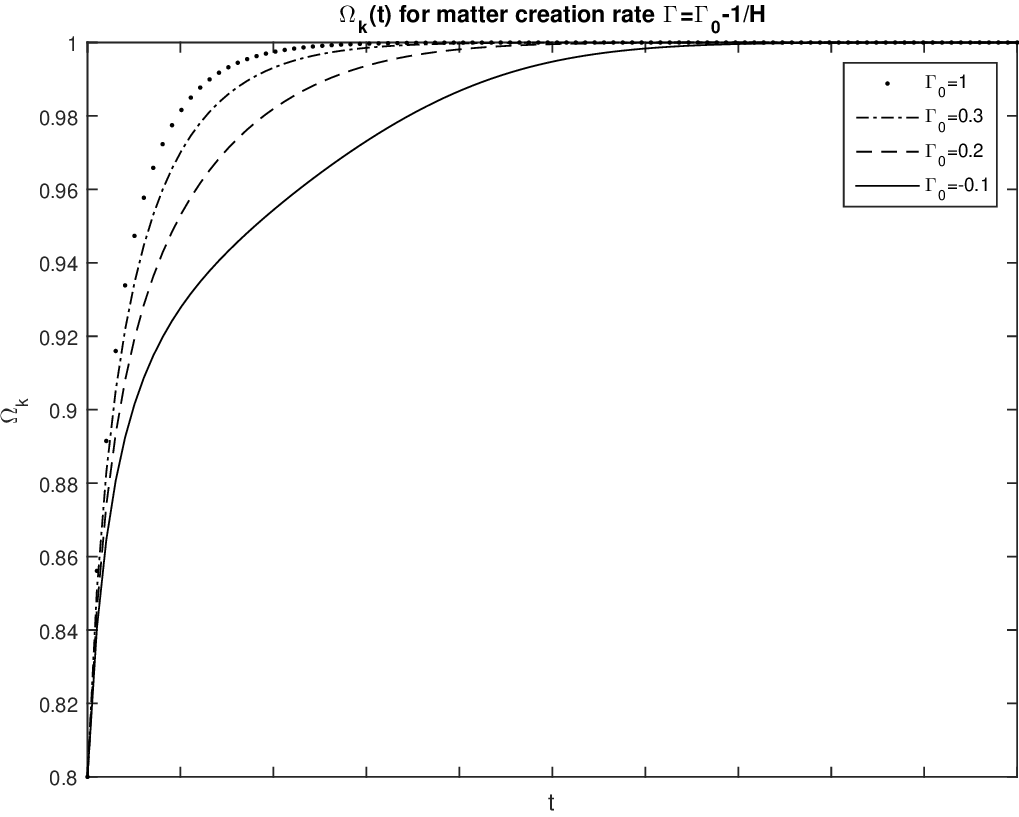}
\caption{Evolution of the energy density $\Omega$ (left figs.) and $\Omega
_{k}~$(right figs.) for the constant matter creation rate $\Gamma_{A}$. Upper
figures are for $\Gamma_{1}=1$ and lower figures are for $\Gamma_{1}=-1$. The
lines are for $\gamma=1$, i.e. dust fluid. Solid lines are for $\Gamma_{0}=1,$
\ dash-dash lines for $\Gamma_{0}=0,$ dot lines for $\Gamma_{0}=-1$ and
dash-dot lines for $\Gamma_{0}=-2.$}%
\label{omegaga1}%
\end{figure}

At the critical points we have $\Omega\left(  P\right)  =\frac{1}{2}$. This
means that $\Omega_{k}\equiv-\frac{k}{a^{2}H^{2}}=\frac{1}{2}$. This is a very
large value, and in disagreement with the cosmological observations for the
post-inflationary epoch at late times. The de Sitter points that were found
above describe the inflationary era, while the field equations (\ref{eq.01}),
(\ref{eq.02}) will describe the evolution of the universe in the
pre-inflationary era. However, in order for the universe to exit from the
inflationary period, new terms in the particle creation rate must dominate the
field equations in order for the de Sitter phase to end. Recall that the de
Sitter points exist only for negative spatial curvature and that is in
agreement with the cosmic scenario, without matter creation, where positive
spatial curvature cannot support inflation \cite{jbnature}.

In Figs. \ref{omegag01} and \ref{omegaga1}, the evolution of $\Omega\left(
t\right)  $ and $\Omega_{k}\left(  t\right)  $ are given for the constant
creation rate, $\Gamma_{0}$, and for $\Gamma_{A}\left(  H\right)  $,
respectively. The plots are for the same initial conditions and $\Omega\left(
t_{0}\right)  =0.2$. We observe that there are two possible late-time
evolutions: the de Sitter point that was derived above, or the solution in
which $\Omega=0$ and $\Omega_{k}=1,$ which corresponds to the ($k=-1$) Milne
universe. When we start with a cosmology different from that of the Milne
universe, we observe that the solution $a\left(  t\right)  =t$, is an
attractor (unstable) and the curvature can dominate in the pre-inflationary epoch.

\begin{figure}[t]
\centering\includegraphics[scale=0.45]{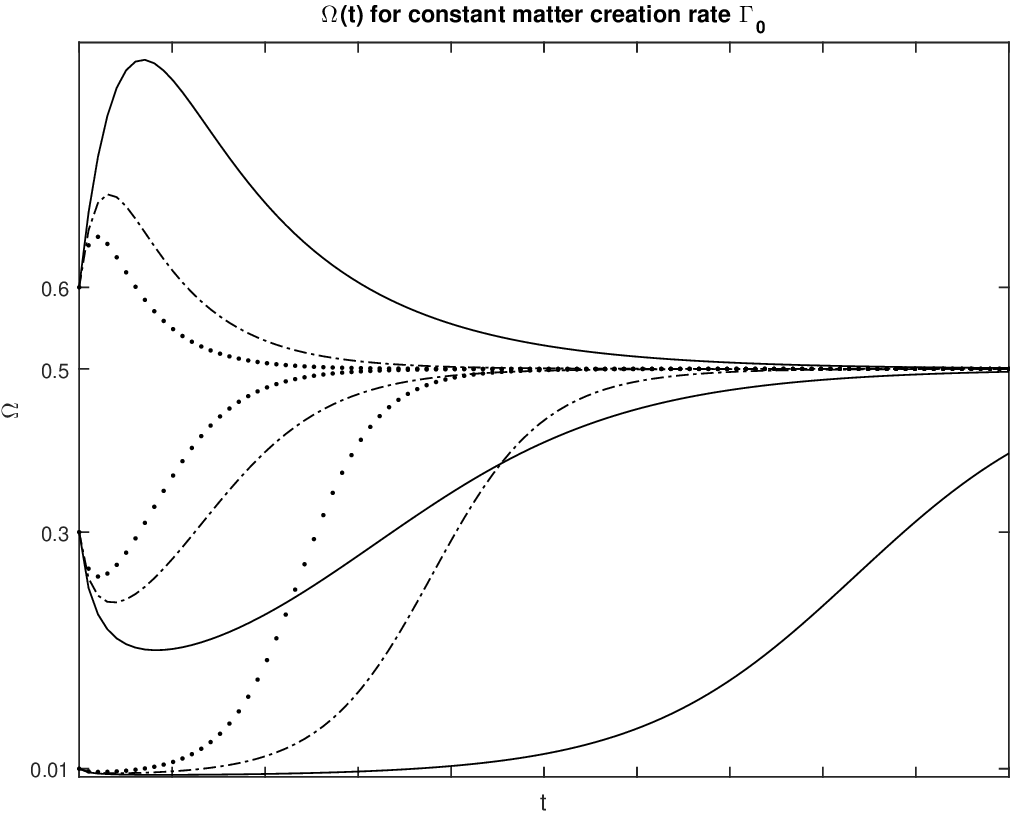}
\centering\includegraphics[scale=0.45]{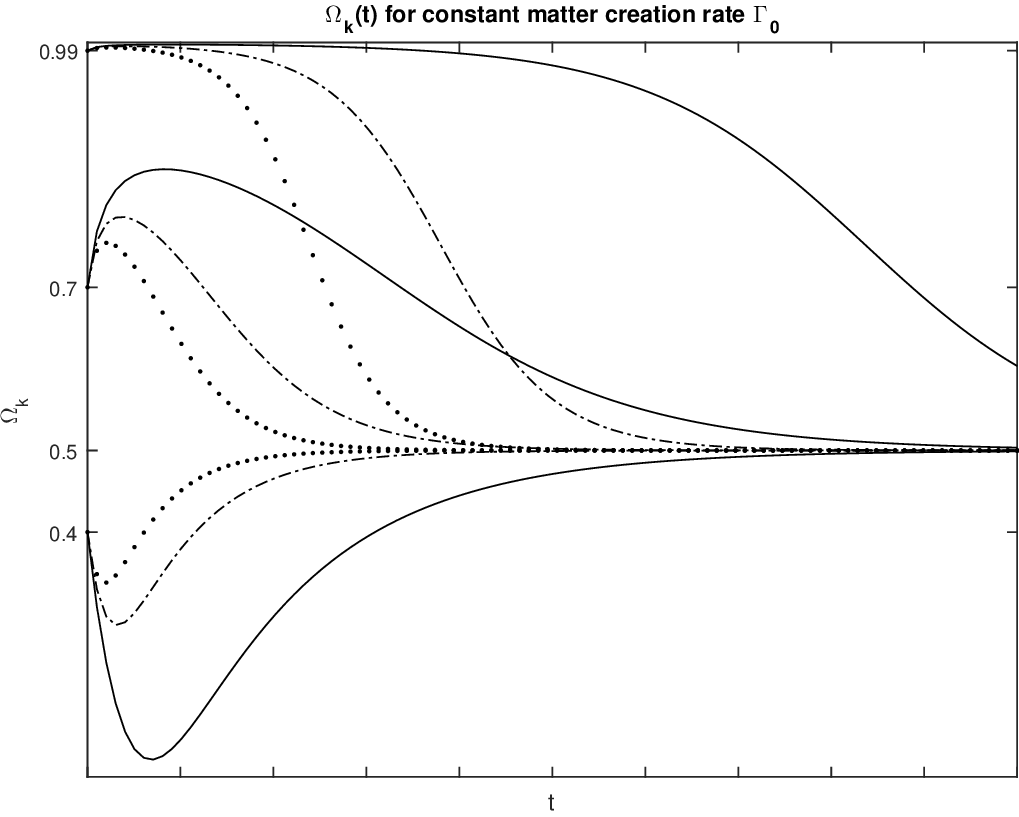}
\caption{Evolution of the energy density $\Omega$ (left fig.) and $\Omega
_{k}~$(right fig.) for the constant matter creation rate $\Gamma_{0}~$and
negative spatial curvature, $k=-1$. The lines are for $\gamma=1$, i.e. dust
fluid and for different initial condition $\Omega\left(  t_{0}\right)  .$
Solid lines are for $\Gamma_{0}=1,$ \ dash-dot lines for $\Gamma_{0}=2,$
dot-dot lines for $\Gamma_{0}=3$.}%
\label{fig33}%
\end{figure}

Furthermore, in Fig. \ref{fig33}, we present the evolution of $\Omega\left(
t\right)  $ and $\Omega_{k}\left(  t\right)  $ for the constant creation rate
for different initial conditions. Specifically, for the initial condition
where the curvature dominates and we are close to the Milne universe (i.e.
$\Omega=0.01$)$,$ this case has $\Omega=0.3$ and $\Omega=0.6,$ which means
that the ideal gas dominates over the curvature. In the three cases for
positive creation rate, $\Gamma_{0}$, the final state of the system is the de
Sitter point.

The dynamical analysis provides us with important information about the
evolution of the system, but to complete our study it is important to see what
is happening to the evolution of the universe before it reaches the de Sitter
points. We proceed to determine the analytical solutions for the master
equation (\ref{ode}) for the three particle creation rates in our study and
also we study the stability of the Milne solution.

\section{Constant creation rate: $\Gamma=\Gamma_{0}$}

\label{sec:constantRate}

Equation (\ref{ode}) can be reduced to a first-order ordinary differential
equation which follows from the autonomy of equation (\ref{ode}), that is,
equivalent to the invariance of the theory under different lapse
transformations of the form, $dt\rightarrow N\left(  \tau\right)  d\tau.$
However, because of the nonlinearity of the theory we apply the method of
singularity analysis to determine the analytic solutions. \

\subsection{Analytic solution}

Let us apply the ARS algorithm \cite{Abl1,Abl2,Abl3} in order to see if
equation (\ref{ode}) passes the singularity test and the solution can be
written as a Laurent expansion. We change variables $a\left(  t\right)
\rightarrow\left(  b\left(  t\right)  \right)  ^{-1}$, so eq. (\ref{ode})
becomes%
\begin{equation}
\left(  b^{2}\right)  ^{^{\prime}}b^{\prime\prime}-\left(  2+3\gamma\right)
\left(  b^{\prime}\right)  ^{3}-\Gamma_{0}\gamma b\left(  b^{\prime}\right)
^{2}+\bar{k}\left(  3\gamma-2\right)  b^{4}b^{\prime}+\Gamma_{0}\gamma\bar
{k}b^{5}=0,\label{sol2}%
\end{equation}
in which $\bar{k}=-k$. \ We see that now eq. (\ref{sol1}) becomes $b\left(
t\right)  =\left(  \bar{k}\right)  ^{-\frac{1}{2}}t^{-1}$. Assume that the
latter has the dominant behavior around the movable singularity $t=z-z_{0}$,
in the complex plane. We assume a solution of the form $b\left(  t\right)
=\left(  \bar{k}\right)  ^{-\frac{1}{2}}t^{-1}+mt^{-1+s}$ in eq. (\ref{sol2})
and linearize it around $m=0$, in order to determine the position of the
resonances. We find that $s=s_{1}=-1$, or $s=s_{2}=2-3\gamma$. The existence
of the first resonance tells us that our analysis is correct and the movable
singularity exists. The second resonance gives us the position of the second
integration constant, where $\gamma\neq1$. Recall that one integration
constant is the position of the singularity $z_{0}$. We see that for $s_{2}%
>0$, we have $\gamma<\frac{2}{3}\,\,$, while $s_{2}<0$ for $\gamma>\frac{2}%
{3}$ and $s_{2}=0$ for $\gamma=\frac{2}{3}$. Hence, for $\gamma\geq\frac{2}%
{3}$, the solution will be given as a right Painlev\'{e} series, while for
$\gamma<\frac{2}{3}$ the solution is expressed in a left Painlev\'{e} series.
\ However, in order to claim that Eq. (\ref{sol2}) passes the singularity
test, \ we need to check the consistency of the solution. To do that let us
write the solution for some values of the barotropic parameter $\gamma$.

Assume that the perfect fluid is radiation-like, i.e. $\gamma=\frac{4}{3}$,
then $s_{2}=-2$, which gives that the Laurent expansion is a left Painlev\'{e}
series; specifically%
\begin{equation}
b\left(  t\right)  =\frac{1}{\sqrt{\left(  \bar{k}\right)  }}t^{-1}%
+b_{2}t^{-2}+b_{3}t^{-3}+%
{\displaystyle\sum\limits_{i=4}^{\infty}}
b_{i}t^{-i},\label{sol3}%
\end{equation}
where, by substituting Eq. (\ref{sol3}) into Eq. (\ref{sol2}), we find that
$b_{2}$ is an arbitrary parameter; that is,\ it is the second integration
constant, while for the rest of the coefficients we have%
\begin{equation}
b_{3}=\left(  b_{2}\right)  ^{2}\left(  \bar{k}\right)  ^{1/2}~\ ,~b_{4}%
=\left(  b_{2}\right)  ^{3}\bar{k}~,~b_{I}=\left(  b_{2}\right)  ^{I-1}\left(
\bar{k}\right)  ^{\frac{I-1}{2}}.\label{sol4}%
\end{equation}
Hence, Eq. (\ref{sol3}) is given by the simple series form%
\[
\frac{b\left(  t\right)  }{b_{2}}=%
{\displaystyle\sum\limits_{i=1}^{\infty}}
\left(  b_{2}\sqrt{\bar{k}}\right)  ^{i-2}t^{-i},
\]
or equivalently $\frac{b\left(  t\right)  }{b_{2}}=\left(  1-\left(  b_{2}%
\bar{k}^{\frac{1}{2}}\,t\right)  ^{-1}\right)  ^{-1}-1,~\,$that is, $a\left(
t\right)  =\sqrt{\bar{k}}\,\,t-\frac{1}{b_{2}}~,$which is also a general
solution for arbitrary $\gamma$.

We now assume the second case where~$\gamma=\frac{1}{2}$, $s_{2}=\frac{1}{2}$,
which means that the Laurent expansion is given by the right Painlev\'{e}
series%
\begin{equation}
b\left(  t\right)  =\left(  \bar{k}\right)  ^{-1}\,t^{-1}+\bar{b}_{2}%
t^{-\frac{1}{2}}+\bar{b}_{3}+\bar{b}_{4}t^{\frac{1}{2}}+%
{\displaystyle\sum\limits_{i=5}^{\infty}}
b_{5}t^{\frac{i}{2}-1}.\label{sol6}%
\end{equation}

Hence, from equations (\ref{sol6}) and (\ref{sol2}), we find that the first
coefficients are
\begin{equation}
\bar{b}_{3}=\frac{19}{16}\left(  \bar{b}_{2}\right)  ^{2}\sqrt{\bar{k}}%
~,~\bar{b}_{4}=\frac{3}{2}\bar{b}_{2}\Gamma_{0}+\frac{47}{32}\left(  \bar
{b}_{2}\right)  ^{3}\bar{k}~,
\end{equation}%
\begin{equation}
\bar{b}_{5}=\frac{31}{40}\left(  \bar{b}_{2}\right)  ^{2}\Gamma_{0}\sqrt
{\bar{k}}+\frac{473}{256}\left(  b_{2}\right)  ^{4}\left(  \bar{k}\right)
^{\frac{3}{2}},~\bar{b}_{6}=...,
\end{equation}
where $\bar{b}_{2}$ is again the second integration constant.\ \ Finally, we
conclude that for $\gamma\neq1$, and $\gamma,\ $a rational number, equation
(\ref{ode}) passes the singularity test. \ The case $\gamma=1$, is of special
consideration since the two resonances are equal and $s_{1}=s_{2}=-1$. In the
case of a radiation-like fluid, i.e. $\gamma=\frac{4}{3}$ and dust-like, i.e.
$\gamma=1$, we do not find any new solution from the singularity analysis
except the well known solution in Eq. (\ref{sol1}). However, as we saw for
other values of $\gamma$, new families of solutions exist.

In order to study the evolution of the universe we continue our analysis by
studying the behaviour of equation (\ref{ode}) at the limits in which the
scale factor is small or large.

\subsection{Solutions at the limits}

\label{Gamma0-limits}

We discussed above that equation (\ref{ode}) can be written as a first-order
ODE. To effect this we transform the time coordinate,~$dt=N\left(
\tau\right)  d\tau$, to obtain%
\begin{equation}
\frac{2}{N^{3}}a^{\prime}a^{\prime\prime}-\frac{2}{N^{4}}a^{\prime}N^{\prime
}+(3\gamma-2)\,\frac{1}{N}\frac{a^{\prime}}{a}\,\left(  \frac{1}{N^{2}}\left(
a^{\prime}\right)  ^{2}+k\right)  -\Gamma_{0}\,\gamma\,\left(  \frac{1}{N^{2}%
}\left(  a^{\prime}\right)  ^{2}+k\right)  =0, \label{sol.13}%
\end{equation}
where prime `$^{\prime}$' denotes the derivative with respect to $\tau$. We
see that when $N\left(  t\right)  =const,$ then $t=N_{0}\tau$. Recall also
that in the new coordinates the Hubble function is $H\left(  \tau\right)
=\frac{1}{N\left(  \tau\right)  }\frac{a^{\prime}}{a}.$

Without any loss of generality pick $N\left(  \tau\right)  $ so that $a\left(
\tau\right)  $ is a linear function, i.e. $a\left(  \tau\right)  =\tau$.
Hence, equation (\ref{sol.13}) becomes the first-order ODE with dependent
variable $N\left(  a\right)  ,$%

\begin{equation}
2\frac{dN}{da}-\left[  \,\frac{(3\gamma-2)}{a}-\Gamma_{0}\,\gamma\,N\right]
N\left(  1+kN^{2}\right)  =0.\label{sol.14}%
\end{equation}
Mathematically it is possible to perform the reduction from a second-order to
a first-order ODE\ \ because the original equation (\ref{ode}) is autonomous.
\ But Eq. (\ref{sol.14}) is non-separable and its solution cannot be found by
quadratures. Yet, we see that for $\left(  1+kN^{2}\right)  =0$, i.e.
$N=1/\sqrt{-k}$, we have the solution (\ref{sol1}). Furthermore, we can write
the solution of eq. (\ref{sol.14}) in the limits $a\rightarrow0$ and
$a\rightarrow\infty.$ This reveals the dominant term in the original system.
\ Of course, eq. (\ref{ode}) can be written as a first-order ODE in terms of
the Hubble function $H\left(  t\right)  $; but the main difference is that we
selected the independent variable to be the scale-factor factor $a$, which
provides the evolution of the Hubble function $H\left(  a\right)  =\left(
aN\left(  a\right)  \right)  ^{-1}$ in terms of the various scales (or
redshifts) of the universe. Since any differentiable non-constant function is
locally monotonic, locally there exists an expression in terms of the
\textquotedblleft time parameter $\tau$ \textquotedblright\ or the proper time
$t$, (for which $N\left(  t\right)  =const$.).\ \ Moreover, with that
representation we transfer the problem from evolution in time to evolution
with respect to the scale factor of the universe.

Hence, for very large scale factor, that is, when $a\rightarrow\infty$,
equation (\ref{sol.14}) takes the simple form%
\begin{equation}
2\frac{dN}{da}+\Gamma_{0}\,\gamma\,N^{2}\left(  1+kN^{2}\right)
\simeq0,\label{sol.15}%
\end{equation}
which gives
\begin{equation}
-\frac{2}{\Gamma_{0}\gamma}\int\left(  N^{2}\left(  1+kN^{2}\right)  \right)
^{-1}dN\simeq a~,~a\rightarrow\infty;
\end{equation}
and solution%
\begin{equation}
\frac{1}{N}+\arctan\left(  N\right)  \simeq\frac{\Gamma_{0}\gamma}{2}a~,~k=1,
\end{equation}
or%
\begin{equation}
\frac{1}{N}-\frac{1}{2}\ln\left(  \frac{N+1}{N-1}\right)  \simeq\frac
{\Gamma_{0}\gamma}{2}a~,~k=-1,\label{ss10}%
\end{equation}
from which we can see that for $k=-1$, as $a\rightarrow\infty,$ it follows
that $N\rightarrow1,$ which satisfies the constraint eq. (\ref{friedmann1})
and $\rho=0.$

Alternatively, to study the behavior in the very early universe, i.e. when
$a\rightarrow0$, we perform the change of variable $a\rightarrow\frac{1}{T}$
in equation (\ref{sol.14}) and find%
\begin{equation}
\frac{dN}{dT}=\frac{\left(  3\gamma-2\right)  \left(  1+kN^{2}\right)  }%
{2T}-\frac{\Gamma_{0}\gamma N^{2}\left(  1+kN^{2}\right)  }{2T^{2}}.
\label{sol.15a}%
\end{equation}
Hence, as $T\rightarrow\infty,$ we find $N$ is a constant but keeping the
first correction gives%
\begin{equation}
\frac{dN}{dT}\simeq\frac{\left(  3\gamma-2\right)  \left(  1+kN^{2}\right)
}{2T}, \label{sol.16}%
\end{equation}
and for $3\gamma-2>0$ we find that
\[
H\left(  a\right)  =c_{1}a^{3\gamma-2}-ka^{-1}%
\]
$\mathbf{~}$

\subsection{Stability of the solution $a\left(  t\right)  \simeq t$}

In order to study the stability of the solution $a\left(  t\right)  =t$, for
$k=-1$, when $\Gamma=\Gamma_{0}$, we substitute $a\left(  t\right)
=t+\varepsilon A\left(  t\right)  $ in equation (\ref{ode}) and we linearize
it around $\varepsilon=0$. Under linearization equation (\ref{ode}) becomes%
\begin{equation}
\frac{d^{2}A}{dt^{2}}=\left(  \frac{2-2\gamma}{t}+\gamma\Gamma_{0}\right)
\frac{dA}{dt},
\end{equation}
for which the closed form solution is\ an incomplete gamma function $A\left(
t\right)  \simeq\Gamma_{in}\left(  3-2\gamma,-\gamma\Gamma_{0}t\right)  $, so
as $t\rightarrow\infty$:
\begin{equation}
A\left(  t\right)  \simeq\left(  t\right)  ^{2-2\gamma}\exp\left(
\gamma\Gamma_{0}t\right)  .
\end{equation}

Therefore, $\lim_{t\rightarrow+\infty}A\left(  t\right)  \rightarrow0$, when
$\gamma\Gamma_{0}<0$, which means that the solution $a\left(  t\right)  =t$,
is stable for negative constant matter creation rate and unstable for positive
constant matter creation rate.

For $k=1$, in which the signature of the spacetime (\ref{cp.01}) is Euclidian,
the stability of the solution $a\left(  t\right)  \simeq t$, depends on the
matter creation rate $\Gamma_{0}$ as in the case of the hyperbolic
three-dimensional spacetime. This is an expected result since for negative
matter creation rate, at the end, the universe will be dominated only by the
curvature term i.e. without any extra matter source.

\begin{figure}[t]
\centering\includegraphics[scale=0.50]{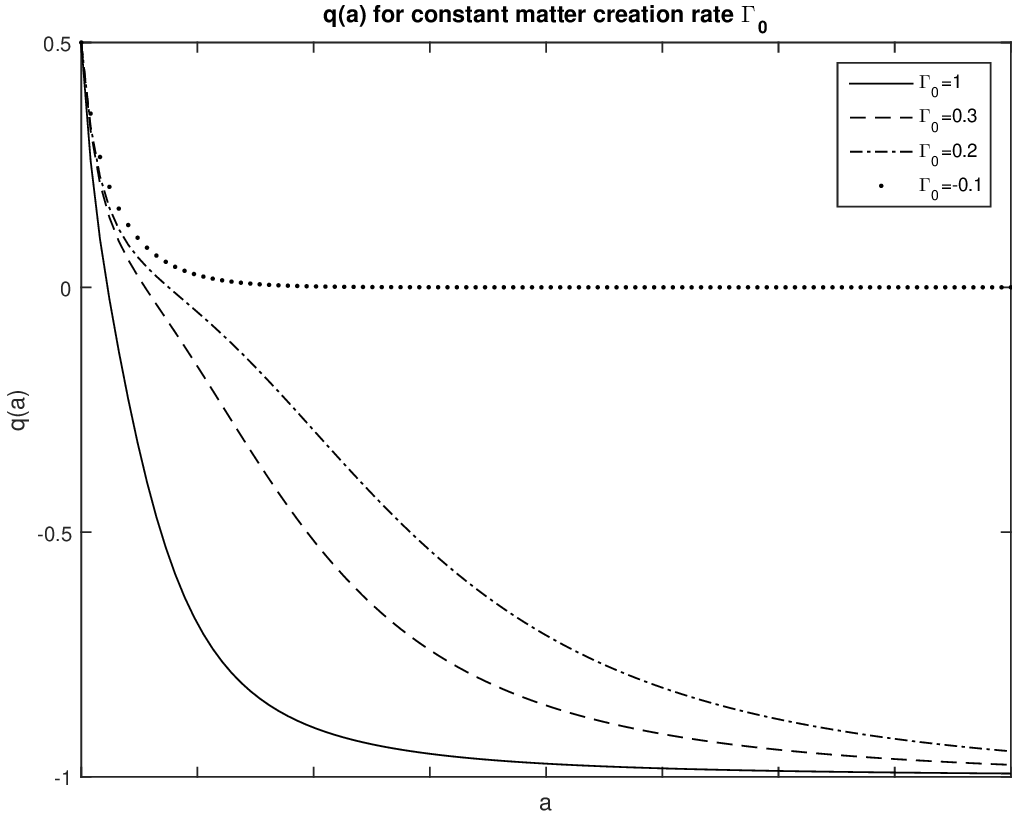}
\caption{Qualitative behaviour of the deceleration parameter parameter
$q\left(  a\right)  $ for the constant matter creation rate $\Gamma_{0}~$and
negative spatial curvature, $k=-1$. The lines are for $\gamma=1$, i.e. dust
fluid. Solid line is for $\Gamma_{0}=1,$ \ dash-dash line for $\Gamma
_{0}=0.3,$ dash-dot line for $\Gamma_{0}=0.2$ and dot-dot line for $\Gamma
_{0}=-0.1.$}%
\label{fig01}%
\end{figure}

The qualitative evolution of the deceleration parameter\footnote{Recall we
have that $q\left(  t\right)  =\frac{\left(  1+3w_{tot}\right)  }{2}\left(
1+\frac{k}{\left(  aH\right)  ^{2}}\right)  ;$ that is, $q\left(  a\right)
=\frac{\left(  1+3w_{tot}\right)  }{2}\left(  1+kN\left(  a\right)
^{2}\right)  .$}, $q\left(  a\right)  ,$ for the constant matter creation rate
is given in Fig. \ref{fig01} for $\gamma=1,~k=-1$, and for various values of
$\Gamma_{0}$. For $\Gamma_{0}>0$, the total fluid has the deceleration
parameter $q\left(  a\right)  =-1$, while for $\Gamma_{0}<0$ the universe is
dominated by the curvature term as expected, and $q=0$, which corresponds to
$w_{tot}=-1/3$. This is in agreement with the results of Section \ref{appendi}
where we found that, for $\Gamma_{0}>0,$ the universe reaches the de Sitter
point. ~

\section{Creation rate: $\Gamma\propto1/H~$and $\Gamma\propto H^{2}$}

\label{sec:varyingI}

We continue our analysis by considering that the particle creation rate is
time varying and of the form $\Gamma_{A}\left(  H\right)  =\Gamma_{0}%
+\Gamma_{1}/H$ or $\Gamma_{B}\left(  H\right)  =\Gamma_{0}+\Gamma_{1}H^{2}$.

\subsection{Analytic Solution}

As in the case of constant matter creation rate in order to study the
existence of analytical solutions we apply the ARS algorithm. We omit the
details and we say that for the model $\Gamma_{B}\left(  H\right)  $ the
second-order differential equation (\ref{ode}) fails to pass the singularity
test. However for $\Gamma_{A}$ the ARS algorithm succeed and gives that the
resonances, the position of the integration constant, are $s_{1}%
=-1~,~s_{2}=2-3\gamma$, where again the dominant behaviour is $a^{-1}=b\left(
t\right)  =\left(  \bar{k}\right)  ^{\frac{1}{2}}t^{-1}$.

In order to compare the results with before we consider $\gamma=\frac{1}{2},$
in which the solution is expressed in terms of the Painlev\'{e} series
(\ref{sol6}), where again $\bar{b}_{2}$ is the second integration constant,
while the first coefficients become%
\begin{equation}
\bar{b}_{3}=\frac{19}{16}\sqrt{\bar{k}}\left(  b_{2}\right)  ^{2}~,~\bar
{b}_{4}=\frac{3}{10}\Gamma_{0}\bar{b}_{2}+\frac{47}{33}\bar{k}\left(  \bar
{b}_{2}\right)  ^{3},
\end{equation}
and%
\begin{equation}
\bar{b}_{5}=\frac{31}{40}\Gamma_{0}\sqrt{\bar{k}}\left(  \bar{b}_{2}\right)
^{2}+\frac{473}{256}\left(  \bar{b}_{2}\right)  ^{4}\left(  \bar{k}\right)
^{\frac{3}{2}},~\bar{b}_{6}=\bar{b}_{6}\left(  \bar{b}_{2},\bar{k},\Gamma
_{0},\Gamma_{1}\right)  .
\end{equation}
from where we can see that first coefficients of the Laurent expansion are
similar to that for $\Gamma\left(  H\right)  =\Gamma_{0}$, which means that
close to the singularity, the $\Gamma_{0}$ term dominates the matter creation
rate. We continue with the study of the solution at the limits.

\subsection{Solutions at the limits}

\subsubsection{Creation rate $\Gamma_{A}$}

As in the case of constant creation rate we consider the transformation
$dt=N\left(  \tau\right)  d\tau$ and the master equation (\ref{ode}) where,
without any loss of generality, we select $a=\tau$, so the master equation
(\ref{ode}) becomes%
\begin{equation}
-2\frac{dN}{da}+(3\gamma-2)\,\frac{N}{a}\left(  1+kN^{2}\right)  -\left(
\Gamma_{0}\,\gamma+\Gamma_{1}\gamma Na\right)  N^{2}\left(  1+kN^{2}\right)
=0. \label{ss.02}%
\end{equation}

As before in the limit $a\rightarrow\infty$ for nonzero $\Gamma_{1}$ we have,
for for nonzero $N$,
\begin{equation}
2\frac{dN}{da}\simeq-\Gamma_{1}\gamma aN^{3}\left(  1+kN^{2}\right)  .
\end{equation}
where the solution of the latter equation is%
\begin{equation}
-\frac{1}{\gamma\Gamma_{1}}\int\frac{dN}{N^{3}\left(  1+kN^{2}\right)  }\simeq
a^{2}.
\end{equation}
which gives%
\begin{equation}
\frac{1}{N^{2}}+k\ln\left(  \frac{N^{2}}{k+N^{2}}\right)  \simeq2\Gamma
_{1}\gamma~a^{2}~,~k=\pm1
\end{equation}

On the other hand, in the limit $a\rightarrow0$, equation (\ref{ss.02})
reduces to that of expression in eq. (\ref{sol.15a}). That result tell us that
in the early universe, when $a\rightarrow0$, only the constant term in the
particle creation rate $\Gamma_{A}$ affects the field equations. That was
observed and before from the singularity analysis

\subsubsection{Creation rate $\Gamma_{B}$}

For the second creation rate $\Gamma_{B}$, in the lapse time $dt=N\left(
\tau\right)  dt$, where $a=\tau$, the master equation becomes%

\begin{equation}
2\frac{dN}{da}=\frac{(3\gamma-2)}{a}\,N\,\left(  1+kN^{2}\right)
-\gamma\left(  \Gamma_{0}+\Gamma_{1}N^{-2}a^{-2}\right)  \,N^{2}\left(
1+kN^{2}\right)  , \label{req.02}%
\end{equation}

In the limit $\tau\rightarrow\infty$, which means~$a\rightarrow\infty$, it
follows from eq. (\ref{req.02}) that the dominant term is that of the constant
creation rate, $\Gamma_{0}$, while if $\Gamma_{0}=0$, we have $N^{\prime
}\simeq0$, hence $N\simeq N_{0}$. The latter means that $\Gamma\left(
\frac{1}{Na}\right)  $ $\rightarrow0$, and for large scale factor values the
matter creation rate does not contribute to the dynamics of the universe.
Furthermore, from condition (\ref{friedmann1}) we find that $\rho\rightarrow
0$, because $aN\rightarrow\infty$.

In the limit of the early universe, that is, $a\rightarrow0$ again we define
the new variable $a\rightarrow T^{-1}$, where now equation (\ref{req.02}) in
the limit $T\rightarrow\infty$, becomes \
\begin{equation}
2\frac{dN}{da}=\gamma\Gamma_{1}\left(  1+kN^{2}\right)  ,
\end{equation}
which gives%
\begin{equation}
N\left(  T\right)  \simeq\tan\left(  \frac{\Gamma_{1}}{2}\gamma T+\phi
_{1}\right)  ~\text{, ~}k=1,
\end{equation}%
\begin{equation}
N\left(  T\right)  \simeq\text{\textrm{tanh}}\left(  \frac{\Gamma_{1}\gamma
}{2}T+\phi_{1}\right)  ~,~k=-1.
\end{equation}
For very large values of~$T$ and for $k=-1$, we find that $N\left(  T\right)
=1$. \ Hence, the solution is that of the vacuum, i.e. $a\left(  t\right)
=t$. We now consider the stability of that special solution.

\subsection{Stability of the solution $a\left(  t\right)  \simeq t$}

In order to study the stability of the solution $a\left(  t\right)  =t$, we
substitute $a\left(  t\right)  =t+\varepsilon A\left(  t\right)  $ in Eq.
(\ref{ode}) and we linearize it around $\varepsilon=0$. For $k=-1$, the
linearized equation for $\Gamma_{A}$ becomes
\begin{equation}
\frac{d^{2}A}{dt^{2}}=\left(  \frac{2-2\gamma}{t}+\gamma\left(  \Gamma
_{0}+\Gamma_{1}t\right)  \right)  \frac{dA}{dt}, \label{sta02}%
\end{equation}
so,%
\begin{equation}
A\left(  t\right)  =\int\exp\left(  \gamma\left(  \Gamma_{0}t+\frac{1}%
{2}\Gamma_{1}t^{2}\right)  \right)  t^{2-2\gamma}dt. \label{sta01}%
\end{equation}

In the limit $\Gamma_{0}\rightarrow0$, $A\left(  t\right)  $ is given by an
incomplete gamma function from which we find that the solution $a\left(
t\right)  =t$ is stable when $\Gamma_{1}<0.$ However, for $\Gamma_{0}\neq0$,
from eq. (\ref{sta02}) as $t\rightarrow0$, we see $A\left(  t\right)  $ is
approximated by the exponential function, $A\left(  t\right)  \simeq
\exp\left(  \Gamma_{1}t^{2}\right)  $, which means that the solution $a\left(
t\right)  =t$, is again stable when $\Gamma_{1}<0$. We note that the same
results hold for $k=1.$

Previously, we found that a negative matter creation rate indicates that the
universe is dominated by the curvature term. However, here we have to mention
that for $\Gamma_{0}>0$ and $\Gamma_{1}<0$, it is possible to have periods in
which the total matter creation rate is positive. But regarding the stability
of the linear solution $a\left(  t\right)  =t$, at large times this solution
will be stable when $\Gamma_{1}<0$.

\begin{figure}[t]
\centering\includegraphics[scale=0.45]{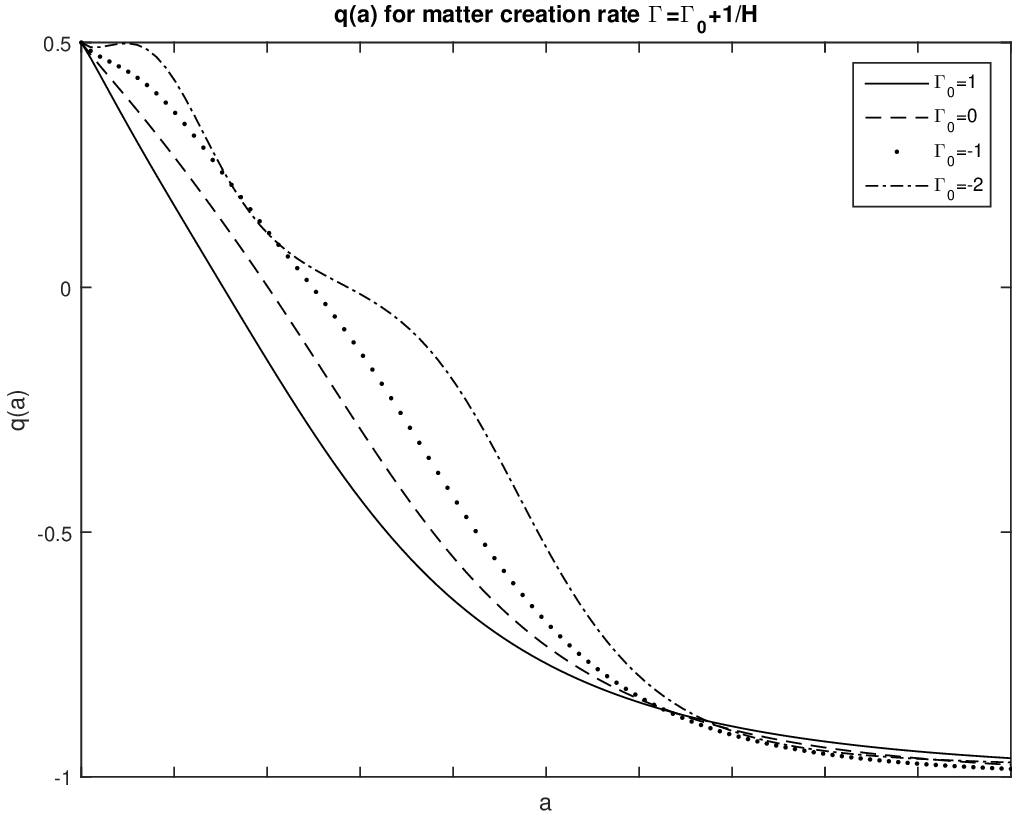}
\centering\includegraphics[scale=0.45]{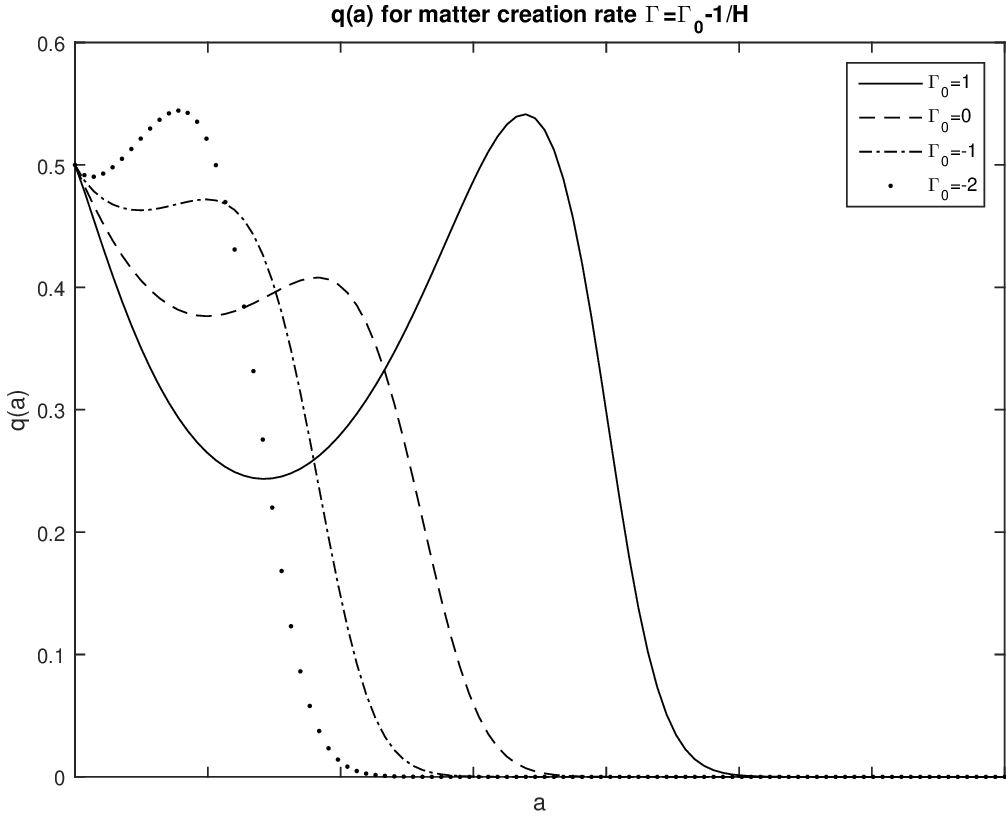}
\caption{Qualitative behaviour of the deceleration parameter parameter
$q\left(  a\right)  $ for the varying matter creation rate $\Gamma\left(
H\right)  =\Gamma_{0}+\Gamma_{1}/H~$and $k=-1$. Left figure is for $\Gamma
_{1}=1$ and right figure is for $\Gamma_{1}=-1$. The lines are for $\gamma=1$,
i.e. dust fluid. Solid lines are for $\Gamma_{0}=1,$ \ dash-dash lines for
$\Gamma_{0}=0,$ dot lines for $\Gamma_{0}=-1$ and dash-dot lines for
$\Gamma_{0}=-2.$}%
\label{fig02a}%
\end{figure}

In Fig. \ref{fig02a} we present the qualitative behavior of the deceleration
parameter $q\left(  a\right)  $ for the matter creation rate $\Gamma\left(
H\right)  =\Gamma_{0}+\frac{\Gamma_{1}}{H}$. We observe that for $\Gamma
_{1}>0$ the total fluid has $q\left(  a\right)  =-1$, while for $\Gamma_{1}<0$
\ the universe is dominated by the curvature term as\ follows from the
stability analysis. The plots in Fig. \ref{fig02a} are made for a dust-like
fluid.

On the other hand as far as concerns the stability of the vacuum solution
$a\left(  t\right)  =t$ for the varying creation rate $\Gamma_{B}$ we find
that the perturbative term around $a\left(  t\right)  =t$, for $k=-1$, is
given to be%
\begin{equation}
A\left(  t\right)  =\int\exp\left(  2\gamma\left(  \Gamma_{0}t-\frac
{\Gamma_{1}}{t}\right)  \right)  t^{4-6\gamma}dt. \label{sta0111}%
\end{equation}

In the limit where $\Gamma_{0}\neq0$, and for $t\rightarrow\infty$ as we have
that solution $a\left(  t\right)  =t$, is stable for $\Gamma_{0}<0$ and
unstable for $\Gamma_{0}>0$. \ However, in the limit $\Gamma_{0}=0$, $A\left(
t\right)  $ is expressed in terms of the Whittaker function and the solution
$a\left(  t\right)  =t$ is stable when $\Gamma_{1}>0$ and unstable when
$\Gamma_{1}<0$.

\section{Time-varying creation rate $\Gamma\left(  H\right)  $}

\label{sec:varyingII}

We now consider a more general function $\Gamma\left(  H\right)  $, and the
transformation $dt=N\left(  \tau\right)  d\tau$, in equation (\ref{ode}).
Without any loss of generality, we again select $a\left(  \tau\right)  =\tau$,
to give the first-order ODE%
\begin{equation}
2\frac{dN}{da}=\frac{(3\gamma-2)}{a}\,N\,\left(  1+kN^{2}\right)
-\gamma\Gamma\left(  \frac{1}{aN}\right)  \,N^{2}\left(  1+kN^{2}\right)  ,
\label{ds.01}%
\end{equation}
Let us study this equation in the limits $a\rightarrow\infty$ and
$a\rightarrow0$.

\subsection{ Limit $a\rightarrow\infty$}

\label{1}

As $a\rightarrow\infty,$ from eq. (\ref{ds.01}) we have
\begin{equation}
2\frac{dN}{da}+\gamma\Gamma\left(  \frac{1}{Na}\right)  \,N^{2}\left(
1+kN^{2}\right)  \simeq0.
\end{equation}

Therefore, we have now to consider special cases for $\Gamma$. If when
$a\rightarrow\infty$, $\Gamma\left(  H\right)  $ is a decreasing function of
$H$, then $N^{\prime}\simeq0$, i.e. $N=N_{0}$. However, if $\Gamma\left(
H\right)  $ is an increasing function in $H$ and we assume that $\Gamma$ is
separable, with $\Gamma=\Phi\left(  a\right)  \Psi\left(  N\right)  ,$ where
$\Phi$, $\Psi$ are any continuous functions in their corresponding variables,
then we find%
\begin{equation}
\int\frac{dN}{\Psi\left(  N\right)  N^{2}\left(  1+kN^{2}\right)  }%
\simeq-\frac{1}{2}\gamma\int\Phi\left(  a\right)  da. \label{sp1}%
\end{equation}
In the special case of $\Gamma\left(  H\right)  =H^{-\alpha}$,~with $\alpha>0$
, we have that $\Gamma\left(  H\right)  =a^{\alpha}N^{\alpha}$; then, solving
Eq. (\ref{sp1}), one has
\begin{equation}
N^{-1-\alpha}L_{\Phi}\left(  -kN^{2},1,-\frac{1+\alpha}{2}\right)  \simeq
\frac{\gamma}{2\left(  \alpha+1\right)  }a^{\alpha+1},
\end{equation}
where $L_{\Phi}\left(  x,c_{1},c_{2}\right)  ~$is the Lerch transcendent
function given by
\begin{equation}
L\left(  x,c_{1},c_{2}\right)  =%
{\displaystyle\sum\limits_{i=0}^{\infty}}
\frac{\left(  x\right)  ^{i}}{\left(  c_{2}+i\right)  ^{c_{1}}}.
\end{equation}
which shows $N^{2}\rightarrow1$ when $a\rightarrow\infty$ and $k=-1$, and
hence $\rho\rightarrow0$.

\subsection{Limit $a\rightarrow0$}

\label{2} \emph{ }

On the other limit, as $a\rightarrow0$, if $\Gamma\left(  H\right)  $ is a
decreasing function in $H$, then the function $\Gamma\left(  H\right)  $ does
not contribute in Eq. (\ref{ds.01}) and it reduces to Eq. (\ref{sol.16}).
Therefore, decreasing terms in $H$ should exist.

Consider the transformation $a\rightarrow T^{-1}$, then equation (\ref{ds.01})
becomes%
\begin{equation}
\frac{dN}{dT}+\frac{(3\gamma-2)}{T}N\,\left(  1+kN^{2}\right)  -\frac{1}%
{T^{2}}\gamma\Gamma\left(  \frac{T}{N}\right)  N^{2}\left(  1+kN^{2}\right)
=0.
\end{equation}
Consider that $\Gamma\left(  \frac{T}{N}\right)  >\frac{1}{T}$ at
$T\rightarrow\infty$, hence $N\left(  T\right)  $ is approximately given by
\begin{equation}
\frac{dN}{dT}-\frac{1}{T^{2}}\gamma\Gamma\left(  \frac{T}{N}\right)
N^{2}\left(  1+kN^{2}\right)  \simeq0,
\end{equation}
and again, if $\Gamma\left(  \frac{T}{N}\right)  =\Theta\left(  T\right)
\Psi\left(  N\right)  $, is a separable function, we find%
\begin{equation}
\int\frac{dN}{\Psi\left(  N\right)  N^{2}\left(  1+kN^{2}\right)  }%
\simeq\gamma\int\frac{\Theta\left(  T\right)  }{T^{2}}dT. \label{ss.0a}%
\end{equation}

Consider the time-varying matter creation rate $\Gamma_{C}\left(  H\right)
=\Gamma_{0}+\frac{\Gamma_{1}}{H}+\Gamma_{2}H^{2}$. As $a\rightarrow\infty$ it
is approximated by $\Gamma\left(  H\right)  \simeq\Gamma_{1}Na$, while as
$a\rightarrow0$, by $\Gamma\left(  H\right)  \simeq\Gamma_{2}\left(
Na\right)  ^{-2}.$ \ As $a\rightarrow\infty,$ the creation rate is dominated
by the $\frac{1}{H}$ function, which we studied above. However, in the early
universe we have $\Psi\left(  N\right)  =N^{-2}$ and $\Theta\left(  T\right)
=T^{2}$. This is exactly the case studied above in section \ref{sec:varyingI}.

\begin{figure}[t]
\centering\includegraphics[scale=0.45]{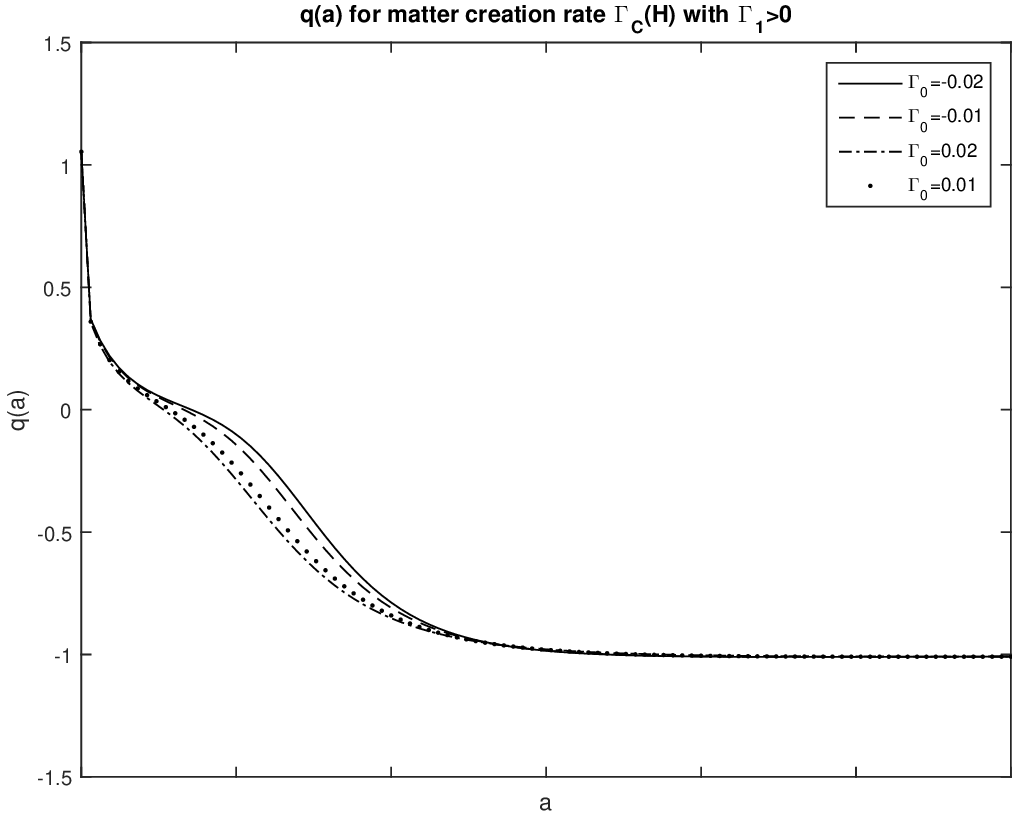}
\centering\includegraphics[scale=0.45]{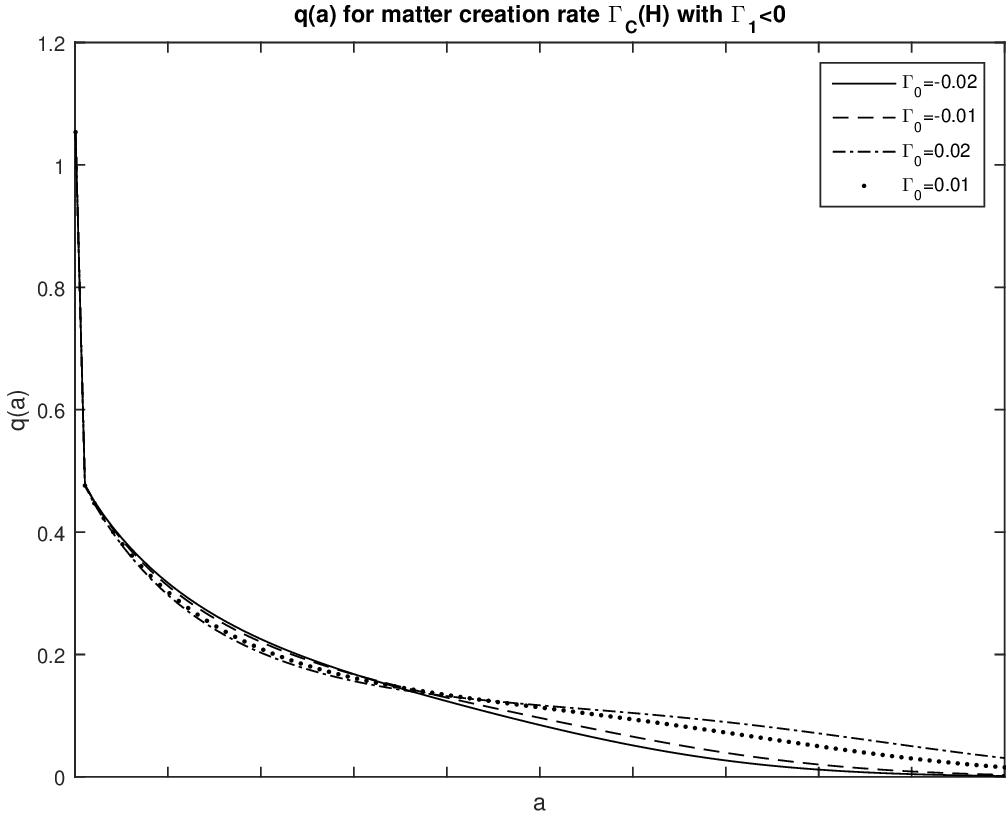}
\caption{Qualitative behaviour of the deceleration parameter parameter
$q\left(  a\right)  $ for the varying matter creation rate $\Gamma\left(
H\right)  =\Gamma_{0}+\Gamma_{1}/H+\Gamma_{2}H^{2}~$and $k=-1$. Left figure is
for $\Gamma_{1}=0.2$ and right figure is for $\Gamma_{1}=-0.2$. The lines are
for $\gamma=1$, i.e. dust fluid. Solid lines are for $\Gamma_{0}=0.2,$
\ dash-dash lines for $\Gamma_{0}=0.1,$ dot lines for $\Gamma_{0}=-0.1$ and
dash-dot lines for $\Gamma_{0}=-0.2.$ For the figures we considered
$\Gamma_{2}=-10^{-4}$. }%
\label{fig04}%
\end{figure}

The qualitative behaviour of $q\left(  a\right)  $ for $\Gamma_{C}\left(
H\right)  $ is given in fig. \ref{fig04} for very small values of the constant
$\Gamma_{2},$ for both $\Gamma_{1}>0$ and $\Gamma_{1}<0$. \ From the figures
we observe that the $H^{2}$ term is significant only at the very early
universe, while the behaviour of the total fluid at late times is independent
of the term $\Gamma_{2}$. However, while the plots are shown for a dust-like
fluid, the term $H^{2}$ introduces a radiation-like fluid which dominates the
very early universe. \

\section{Alternative to curvature effects: A second fluid}

\label{two-fluid}

The possible dominance of the curvature term over the matter creation term at
early times leads us to ask whether this can occur when another perfect fluid
is present, as was found with bulk viscosity present \cite{JDB1988}. Consider
the problem of spatially flat FLRW universe with matter creation and a second
perfect fluid with equation of state parameter { $p_{1}=(\gamma_{1}-1)\rho
_{1}$. If }$\gamma_{1}=\frac{2}{3}$, then this fluid mimics the presence of
spatial curvature in the Friedmann equations and the results for the non-zero
curvature model we considered above are recovered. For the second fluid, the
conservation equation gives\ {$\rho_{1}=\rho_{1,0}\,a^{-3\gamma_{1}}$, where
$\rho_{1,0} >0$, is the present energy density of the second fluid}. Hence, as
before, {the scale factor obeys a second order nonlinear differential
equation:
\begin{equation}
2\dot{a}\ddot{a}+\left(  3\gamma-2\right)  \frac{\left(  \dot{a}\right)  ^{3}%
}{a}-\rho_{1,0}(\gamma-\gamma_{1})a^{1-3\gamma_{1}}\,\dot{a}-\gamma
\Gamma\left(  \dot{a}^{2}-\frac{\rho_{1,0}}{3}a^{2-3\gamma_{1}}\right)  =0.
\label{ode2f}%
\end{equation}
}

We can see that the \ {differential eqn. (\ref{ode}) is recovered from
(\ref{ode2f}) for }$\gamma_{1}=\frac{2}{3}$, and $\rho_{1,0}=-3k$. Moreover,
we observe that eqn. (\ref{ode2f}) admits {the power law solution $a(t)\propto
t^{2/3\gamma_{1}}$,~which describes a universe containing only the fluid
}$\rho_{1}.$ However, as in the case of the curvature this particular solution
need not be stable and its stability depends on the matter creation function,
$\Gamma$.

Regarding the integrability of eqn. (\ref{ode2f}), we can follow our previous
analysis. For $\Gamma=\Gamma_{0}$, when the second fluid is dust ($\gamma
_{1}=1$), we find that the dominate term is $a_{ds}\left(  t\right)
=a_{0}t^{\frac{2}{3}}$, while the resonances are $s_{1}=-1$ and $s_{2}%
=2-2\gamma$. \ Consider the special case in which $\gamma_{1}=\gamma=1$, which
means that both fluids are dust(-like) fluids, but particle creation rate
exists for the fluid defined by $\left(  \rho,p\right)  $. In that special
case, with the use of group invariant transformations, equation (\ref{ode2f})
can be solved with the method of quadratures. For instance, if we select
$\dot{a}\sqrt{a}=w\left(  t\right)  $, then for constant matter creation rate,
we find the solution
\begin{equation}
w\left(  t\right)  =\pm\sqrt{\frac{1}{3}\rho_{1,0}+c_{1}e^{-t}}.
\label{ode2fe}%
\end{equation}

Alternatively, using $z\left(  \tau\right)  =\dot{a}$ and $\tau=a$, we obtain%
\begin{equation}
z\sqrt{\tau}+\frac{2\sqrt{3\rho_{1,0}}}{3}\,\,\text{\textrm{arctanh}}\left(
\frac{\sqrt{3\tau}}{\sqrt{\rho_{1,0}}}z\right)  +\frac{\tau^{\frac{3}{2}}}%
{3}-3c_{1}=0,
\end{equation}
while the reduction $y\left(  \tau\right)  =\dot{a}\sqrt{a}$, $\tau
=\frac{3Ct-2a^{\frac{3}{2}}}{3C}$, yields the solution%
\begin{equation}
C\tau-2y+\ln\left(  3y^{2}-\rho_{1,0}\right)  +\frac{2\sqrt{3\rho_{1,0}}}%
{3}\,\,\text{\textrm{arctanh}}\left(  \frac{\sqrt{3}}{\sqrt{\rho_{1,0}}%
}y\right)  +c_{1}=0,
\end{equation}
where $c_{1}$ and $C$ are constants.

From Eq. (\ref{ode2fe}) we can easily find the scale factor in the closed
form
\begin{equation}
a^{\frac{3}{2}}=\pm\sqrt{3\rho_{1,0}}\left(  \text{\textrm{arctanh}}\left(
\sqrt{1+\frac{3\,c_{1}}{\rho_{1,0}}e^{-t}}\right)  -\sqrt{1+\frac{3\;c_{1}%
}{\rho_{1,0}}e^{-t}}\,\,\right)  +c_{2}, \label{ode2fc}%
\end{equation}
which is only one particular solution of the problem where $c_{2}$ is another
constant. An immediate observation from the above solution of the scale factor
shows that it may encounter with a finite time singularity. For $c_{2}=0$, one
finds that the singularity occurs at $t=t_{f}=\ln\left(  -\frac{3\,c_{1}}%
{\rho_{1,0}}\right)  ~$, therefore for $c_{1}<0$, $t_{f}$ becomes real and the
above solution for the scale factor admits a finite time singularity. Also for
$t\rightarrow\infty$, $a\rightarrow\infty$, that means in future evolution no
finite time singularity is observed. {In the same way, except the constant
creation rate other models can be examined. However,} the analysis of these
models exceeds the purpose of this work and we end it here. \

\section{Conclusions}

\label{discu}

The theory of gravitational particle production has been the main subject of
this paper in situations where there is a non-zero spatial curvature of a
homogeneous and isotropic universe. It is important to mention that the
late-time observations favour spatial curvature of the universe that is very
close to zero \cite{Ade:2015xua, DiDio:2016ykq} and this is one of the natural
consequences of simple inflationary scenarios. However, the appearance of
non-zero spatial curvature does not rule out the inflationary universe. It is
possible to have inflationary cosmologies in noticeably closed universes today
\cite{Linde:2003hc,Bonga:2016iuf} as well as with noticeably open universes
today
\cite{Gott,Kamionkowski:1993cv,Kamionkowski:1994sv,Linde1996,Bucher:1994gb,Linde:1998gs}%
. Now in such non-zero spatial curvature scenarios we gave the gravitational
field equations and reduced them to a non-autonomous second-order differential
equation. In order to prove the existence of solutions, we used the method of
singularity analysis. In particular, we searched for movable singularities for
the matter creation rate functions \ (a) $\Gamma=\Gamma_{0}=$ constant, (b)
$\Gamma_{A}(H)=\Gamma_{0}+\Gamma_{1}/H~$and (c) $\Gamma_{B}\left(  H\right)
=\Gamma_{0}+\Gamma_{1}H^{2}$. We showed that for functions (a) and (b) the
differential equations can pass the singularity test and the analytical
solution can be expressed in terms of Laurent series expansions. The latter is
not true for the creation rate (c).

A special exact solution exists with a linear scale factor $a\left(  t\right)
=t,$ which corresponds to the Milne solution of vacuum spacetime with negative
spatial curvature. We studied the stability of this solution and found that
for $\Gamma_{0}<0$ in model (a), and $\Gamma_{1}<0$ in model (b), the linear
scale factor is a stable solution to the future. For the model (c), the linear
solution is stable for $\Gamma_{0}<0$. That was also demonstrated by the study
of the qualitative behaviour for the total equation of state parameters. We
found that when $\Gamma_{0}>0$ and $\Gamma_{1}>0$ in the two models (a) and
(b), respectively, the total fluid acts as a cosmological constant, while in
case (c), the total fluid has an equation of state parameter like a spatial
curvature, with $w_{tot}=-1/3$.

In addition, we studied the critical points of the system (\ref{friedmann1}),
(\ref{friedmann2}),\ (\ref{conservation}) for the particle creation rates
$\Gamma\left(  H\right)  =\Gamma_{0},~\Gamma\left(  H\right)  =\Gamma_{A}$ and
$\Gamma\left(  H\right)  =\Gamma_{B}$. Every critical point corresponds to a
de Sitter phase but we found that the critical points satisfy the constraint
equation (\ref{friedmann1}) only when the spatial curvature is negative. The
de Sitter point of constant matter-creation rate is an attractor for
$\Gamma_{0}>0$, while the other two points always give an unstable de Sitter
point and an attractive de Sitter point when the constant~$\Gamma_{1}$ is
positive for $\Gamma_{A}$, or negative for $\Gamma_{B}$, respectively.$~$%
However, the analysis of the critical points does not provide us with
information about the behaviour of the solutions when they are not exactly
described by the de Sitter universe.

Furthermore, the qualitative evolution of the cosmological model for a general
time-varying matter-creation function was studied. We found the conditions
under which the universe is dominated by the curvature term or by the matter
creation term. \ An efficient method to study the evolution of the universe is
the analysis of the critical points of the master equation. This reveals the
points which correspond to different phases in the evolution of the universe.
At these points, the solution of the scale factor is a power-law, or an
exponential function of $t$. In the latter case the fixed point describes the
de Sitter universe. This means that we do not know the analytic expression far
from the fixed points, except in the first approximation. In our analysis we
proved the existence of actual solutions to the master equation and then we
found the approximate solutions at the limits.

A striking feature we found is that for non-zero spatial curvature in such a
formalism the Milne solution $a(t)\propto t$, always exists for any particle
creation rate, $\Gamma$; this is not true if particle creation is considered
in the zero-curvature FLRW universe. This is the main difference between the
flat and curved FLRW universes when particle creation is included. These
differences illustrate a feature of FLRW cosmologies with non-linear equations
of state linking the pressure and the density: although the spatial curvature
may be negligible in the universe today it is not necessarily ineligible at
very early times when a non-linear coupling between $p,\rho,$ and $H$ will arise.

For the case with particle creation rate $\Gamma\left(  H\right)  =\Gamma_{0}%
$, we show that the Milne universe is a stable solution for $\Gamma_{0}<0$,
and unstable for $\Gamma_{0}>0$. However, at the de Sitter points, the density
of the curvature term is $\Omega_{k}=\frac{1}{2}$ which is equal to that of
matter source, i.e. $\Omega_{k}=\Omega$. This means that this accelerated
phase should correspond to the inflation era. \ It is important to mention
that if the de Sitter attractor is a stable point then the universe will be
unable to escape from the inflationary phase unless the new particle creation
rate dominates the field equations and changes the stability of the de Sitter
attractor; while at the same time the energy density of the matter source
should dominate the post-inflationary phase in order for $\Omega_{k}~$to be
small as is required by the recent astronomical data.\ Lastly, the stability
conditions for the Milne universe were determined for all other creation
rates, namely $\Gamma_{A}$ and $\Gamma_{B}.$

Finally, we showed how in the case of a spatially-flat FLRW universe the
presence of a second perfect fluid with an appropriate equation of state can
reproduce the effects of curvature. Some special solutions have been given
while the complete analysis of these models is the subject of a forthcoming work.

\begin{acknowledgments}
The authors thank the referee for some important comments which helped to
improve the paper. AP acknowledges the financial support \ of FONDECYT grant
no. 3160121. JDB is supported by the Science and Technology Facilities Council
(STFC) of the United Kingdom. SP is supported by the SERB-NPDF grant (File No:
PDF/2015/000640), Govt. of India.
\end{acknowledgments}


\end{document}